\documentclass[11pt]{article}

\usepackage{epsf,amsmath,amssymb,graphicx,scalefnt,rotating,enumitem,fancyhdr}
\usepackage{array} \usepackage{longtable} \usepackage{slashed}
\usepackage{filemod} \usepackage{multirow}
\usepackage[affil-it]{authblk} \usepackage[usenames,dvipsnames]{color}
\usepackage[pdftex, colorlinks=true, linkcolor=black, filecolor=black,
  citecolor=black, urlcolor=black, draft=false, bookmarks,
  bookmarksnumbered=true, plainpages=false,
  linktocpage={true}]{hyperref}
\usepackage[final]{showlabels}
\showlabels{cite}

\usepackage[normalem]{ulem}
\usepackage[noadjust]{cite}
\usepackage[tocflat]{tocstyle}
\usetocstyle{standard}
\newcounter{notecount}

\newcommand{\gammagf}{\hat{\gamma}^{(\rho)}}
\newcommand{\betagf}[1]{\hat{\beta}_{\rho\ifthenelse{\equal{#1}{}}{}{, }#1}}
\newcommand{\massgf}{\hat{m}^{(\rho)}}
\newcommand{\alphagf}{\hat\alpha_\rho}
\newcommand{\two}{two}
\newcommand{\four}{four}
\newcommand{\three}{three}
\newcommand{\gff}{\abbrev{GFF}}

\newcommand{\TR}{T_\mathrm{R}}
\newcommand{\TF}{T_\mathrm{F}}
\newcommand{\NF}{n_\mathrm{F}}
\newcommand{\CF}{C_\mathrm{F}}
\newcommand{\CA}{C_\mathrm{A}}
\newcommand{\NA}{N_\mathrm{A}}
\newcommand{\NC}{N_\mathrm{c}}

\newcommand{\citere}[1]{Ref.\,\cite{#1}}
\newcommand{\citeres}[1]{Refs.\,\cite{#1}}

\newcommand{\abbrev}[1]{{\scalefont{.9}#1}}
\newcommand{\EulerGamma}{\gamma_\mathrm{E}}
\newcommand{\ep}{\epsilon}
\newcommand{\alphas}{\alpha_\mathrm{s}}

\newcommand{\eqn}[1]{Eq.\,(\ref{#1})}
\newcommand{\noeqn}[1]{(\ref{#1})}
\newcommand{\eqs}[1]{Eqs.\,(\ref{#1})}
\newcommand{\fig}[1]{Fig.\,\ref{#1}}

\newcommand{\sct}[1]{Sect.\,\ref{#1}}
\newcommand{\scts}[1]{Sects.\,\ref{#1}}
\newcommand{\dd}{{\rm d}}
\newcommand{\deriv}[2]{\frac{\partial #1}{\partial #2}}
\newcommand{\dderiv}[2]{\frac{\dd #1}{\dd #2}}
\newcommand{\order}[1]{{\cal O}(#1)}

\newcommand{\qcd}{\abbrev{QCD}}

\newcommand{\pdf}{\abbrev{PDF}}

\newcommand{\nlo}{\abbrev{NLO}}
\newcommand{\nnlo}{\abbrev{NNLO}}

\newcommand{\msbar}{\ensuremath{\overline{\mbox{\abbrev{MS}}}}}

\newcommand{\ii}{\mathrm{i}}

\newcommand{\RHheaderline}{%
  \small\sf
  September 2019\\
  FERMILAB-PUB-19-151-T\\
  FR-PHENO-2019-003\\
  IIT-CAPP-19-02\\
  TTK-19-15
}
\fancypagestyle{firstpage}
{
  
  \fancyhead[R]{\RHheaderline}
}
\title{Results and techniques for higher order calculations within the
  gradient-flow formalism}
\author[1]{Johannes Artz}
\author[1]{Robert V. Harlander}
\author[1]{Fabian Lange}
\author[2,3]{Tobias Neumann}
\author[4]{Mario Prausa}
\affil[1]{Institute for Theoretical Particle Physics and Cosmology,\protect\\
RWTH Aachen University, D-52056 Aachen, Germany}
\affil[2]{Department of Physics, Illinois Institute of Technology,\protect\\
  Chicago, Illinois 60616, USA}
\affil[3]{Fermilab, PO Box 500, Batavia, Illinois 60510, USA}
\affil[4]{Physikalisches Institut, Albert-Ludwigs-Universit\"at, D-79085 Freiburg, Germany}
\begin{document}
\date{}
\maketitle
\thispagestyle{firstpage}
\begin{abstract}
  We describe in detail the implementation of a systematic perturbative
  approach to observables in the \qcd{} gradient-flow formalism.  This
  includes a collection of all relevant Feynman rules of the
  five-dimensional field theory and the composite operators considered
  in this paper. Tools from standard perturbative calculations are used
  to obtain Green's functions at finite flow time $t$ at higher orders
  in perturbation theory. The three-loop results for the quark
  condensate at finite $t$ and the conversion factor for the ``ringed''
  quark fields to the \msbar\ scheme are presented as applications. We
  also re-evaluate an earlier result for the three-loop gluon
  condensate, improving on its accuracy.
\end{abstract}
\newpage
\headheight 0pt
\tableofcontents

\section{Introduction}\label{sec:intro}

The gradient flow in field theory has proven to be a very useful
concept.  Originally introduced as a method to regularize divergences
and ultraviolet fluctuations in lattice
calculations\,\cite{Narayanan:2006rf,Luscher:2009eq}, it has mainly
gained traction through its use for scale-setting in lattice
\qcd{}\,\cite{Luscher:2010iy,Luscher:2013vga,Borsanyi:2012zs} and has
meanwhile become an indispensable tool in many practical calculations in
this field (see,
e.g.\ \citeres{Luscher:2013cpa,Sommer:2014mea,Ramos:2015dla,Aoki:2019cca}).
While first focused on \qcd{}, generalizations of the gradient-flow
formalism (\gff) yield a much wider range of applications, for example
the study of dualities in field
theory\,\cite{Aoki:2015dla,Aoki:2016ohw,Aoki:2017uce,Aoki:2017bru,
  Aoki:2018dmc,Aoki:2019bfb}.

In \qcd{}, the formulation as a five-dimensional field theory has
been presented in \citere{Luscher:2011bx}, where the additional
dimension is associated with the so-called flow time. The form of the
fundamental theory (i.e., actual \qcd{}) serves as the boundary
condition at vanishing flow time. It has been proven that, when
expressed in terms of renormalized parameters and fields, composite
operators at positive flow time are finite\,\cite{Luscher:2011bx}. This
property and the realization that the gradient flow can be used as a
matching scheme between lattice and perturbative calculations shed light
on a whole new set of applications that are currently being explored.

One of the first possible cross-fertilizations among perturbative and
lattice \qcd{} is given by the definition of a new scheme for the strong
coupling, defined by the gluon condensate (\qcd{} action density) at
finite flow time\,\cite{Luscher:2010iy}.  Its perturbative relation to
the \msbar{} coupling is known through three
loops\,\cite{Harlander:2016vzb}.\footnote{Unless stated otherwise, we
  refer to perturbative calculations as performed at infinite volume in
  this paper. The inclusion of finite-volume effects requires different
  techniques, such as Numerical Stochastic Perturbation Theory, see
  \citere{DallaBrida:2017tru}.}  Another cross-boundary application is
the gradient-flow definition of the energy-momentum tensor, which uses
the small-flow-time expansion in order to express it in terms of
well-defined composite operators at positive flow time and
perturbatively accessible coefficient
functions\,\cite{Suzuki:2013gza,Makino:2014taa,
  Harlander:2018zpi,Iritani:2018idk}. Yet another example is the
proposal to relate Euclidean quasi \pdf{}s\footnote{\pdf=parton
  distribution function.} on the lattice to perturbative light-front
\pdf{}s by using the gradient
flow\,\cite{Monahan:2016bvm,Monahan:2017hpu}.

All these applications rely on input from the lattice as well as from
perturbative calculations, both at finite flow time. It was found that
higher order corrections in perturbation theory are crucial for reducing
the perturbative truncation error as estimated by the variation of the
renormalization scale\,\cite{Harlander:2016vzb,Harlander:2018zpi}. Such
corrections were also found to significantly stabilize the required
extrapolations of the corresponding lattice
results\,\cite{Iritani:2018idk}.

Despite the fact that loop integrals in the \gff\ involve additional
exponential factors as well as integrations over flow-time variables,
many important techniques for regular perturbative calculations retain
their usefulness, albeit in slightly generalized form. It is the goal of
this paper to provide the basis for further perturbative calculations
within the \gff, and thus to contribute to the further exploration of
the capabilities of this approach. \sct{sec:pert} reviews the \qcd{}
gradient flow in perturbation theory, recapitulating results of
\citeres{Luscher:2010iy,Luscher:2011bx,Luscher:2013cpa}. The various
stages of automation in a perturbative multi-loop calculation of
correlation functions are described in \sct{sec:implementation}: the
generation of Feynman diagrams and insertion and evaluation of Feynman
rules, followed by a reduction of the loop integrals to a set of master
integrals, and finally the numerical evaluation of the latter. As an
application, \sct{sec:observables} presents the three-loop results for
the gluon condensate, obtained before in \citere{Harlander:2016vzb}, the
quark condensate, as well as the ``ringed'' quark field renormalization
constant, which are new results.  These quantities allow one to define a
precision gradient-flow coupling and gradient-flow mass scheme and their
matching to \msbar{} as shown in \sct{sec:couplingandmass}. Our
conclusions are presented in \sct{sec:conclusions}.

\section{The \qcd\ gradient flow in perturbation theory}
\label{sec:pert}

For completeness, we collect the main steps of the original
derivations of \citeres{Luscher:2010iy,Luscher:2011bx} in this section.

\subsection{Definition of the gradient flow}

In the following, we work in $D$-dimensional Euclidean space-time
with $D=4-2\ep$.  The \gff\ continues the gluon and quark fields
$A^a_\mu(x)$ and $\psi_\alpha^i(x)$ of regular\footnote{We use the terms
  ``flowed'' and ``regular'' \qcd\ to distinguish quantities defined at
  $t>0$ from those defined at $t=0$.} \qcd\ to $(D+1)$-dimensional
fields $B^a_\mu(t,x)$ and $\chi_\alpha^i(t,x$) through the boundary
conditions
\begin{equation}
  \begin{split}
    B_\mu^a (t=0,x) = A_\mu^a (x)\,,\qquad \chi_\alpha^i (t=0,x)=
    \psi_\alpha^i(x)
    \label{eq:bound}
  \end{split}
\end{equation}
and the flow equations\,\cite{Luscher:2010iy,Luscher:2013cpa}
\begin{equation}
  \begin{split}
    \partial_t B_\mu^a &= \mathcal{D}^{ab}_\nu G_{\nu\mu}^b + \kappa
    \mathcal{D}^{ab}_\mu \partial_\nu B_\nu^b\,,\\ \partial_t \chi &= \Delta \chi
    - \kappa \partial_\mu B_\mu^a T^a \chi\,,\\ \partial_t \bar
    \chi &= \bar \chi \overleftarrow \Delta + \kappa \bar \chi
    \partial_\mu B_\mu^a T^a\,,
    \label{eq:flow}
  \end{split}
\end{equation}
where the ``flow time'' $t$ is a parameter of mass dimension minus two,
and $\kappa$ is a gauge parameter which drops out of physical
observables (see below).

The $(D+1)$-dimensional field-strength tensor is defined as
\begin{align}
  G_{\mu\nu}^a = \partial_\mu B_\nu^a - \partial_\nu B_\mu^a + f^{abc}
  B_\mu^b B_\nu^c\,,
\end{align}
the covariant derivative in the adjoint representation is given by
\begin{align}
  \mathcal{D}_\mu^{ab} = \delta^{ab} \partial_\mu - f^{abc} B_\mu^c\,,
\end{align}
and
\begin{equation}
  \begin{split}
    \Delta = \mathcal{D}^\mathrm{F}_\mu \mathcal{D}^\mathrm{F}_\mu
    \,,\qquad
    \overleftarrow{\Delta} = \overleftarrow{\mathcal{D}}^\mathrm{F}_\mu
    \overleftarrow{\mathcal{D}}^\mathrm{F}_\mu \,,
  \end{split}
\end{equation}
with the covariant derivative in the fundamental representation,
\begin{equation}
  \mathcal{D}^\mathrm{F}_\mu = \partial_\mu + B^a_\mu T^a \,,\qquad
  \overleftarrow{\mathcal{D}}^\mathrm{F}_\mu =
    \overleftarrow{\partial}_\mu - B^a_\mu T^a \,.
\end{equation}
As usual, the color indices of the adjoint representation are denoted by
$a,b,c,\ldots$, while $\mu,\nu,\rho,\ldots$ are $D$-dimensional Lorentz
indices. Color indices of the fundamental representation are denoted by
$i,j,k,\ldots$, but they are suppressed throughout this paper, unless
required by clarity; similarly for spinor indices
$\alpha,\beta,\gamma,\ldots$. The symmetry generators $T^a$ are understood in
the fundamental representation. The obey the commutation relation
\begin{equation}
  \begin{split}
    [T^a,T^b] = f^{abc}T^c\,,
  \end{split}
\end{equation}
with the structure constants $f^{abc}$ and the trace is normalized to
\begin{equation}
  \mathrm{Tr}(T^a T^b) = - \TR \delta^{ab} \, ,
\end{equation}
with $\TR > 0$.

Different choices of $\kappa$ in \eqn{eq:flow} correspond to gauge
transformations of the form
\begin{align}
  \chi \rightarrow \chi^\prime = \Lambda \chi \quad \quad \text{and}
  \quad \quad B_\mu^a T^a \rightarrow B_\mu^{a\prime} T^a= \Lambda
  B_\mu^a T^a \Lambda^{-1} + \Lambda \partial_\mu \Lambda^{-1}\,,
\end{align}
where
\begin{align}
  \Lambda(t,x)=e^{-\int_0^t \mathrm{d}s \, \kappa \partial_\mu B_\mu^a(s,x) T^a}\,.
\end{align}
Of course, all observables are independent of the gauge parameter
$\kappa$\,\cite{Luscher:2010iy}.  In perturbative calculations, it is
usually most convenient to set $\kappa = 1$.

The flow equations \noeqn{eq:flow} can be incorporated in a Lagrangian
formalism by defining
\begin{align}
  \mathcal{L} = \mathcal{L}_\mathrm{\qcd} +
  \mathcal{L}_\textrm{gauge-fixing} + \mathcal{L}_\mathrm{ghost} +
  \mathcal{L}_{B} + \mathcal{L}_{\chi}.
  \label{eq:Lagrangian_complete}
\end{align}
The first three terms constitute the regular Yang-Mills Lagrangian, with
fermions added in the fundamental representation (quarks). Introducing an
index $f$ in order to distinguish different quark flavors of mass $m_f$,
the classical, gauge-fixing, and Faddeev-Popov ghost part are given by
\begin{equation}
  \begin{split}
  \mathcal{L}_\mathrm{\qcd} &= \frac{1}{4g^2} F_{\mu \nu}^a F_{\mu
    \nu}^a + \sum_{f=1}^{\NF}\bar{\psi}_f ( \slashed{D}^\mathrm{F}+m_f)
  \psi_f\,,\\ \mathcal{L}_\textrm{gauge-fixing}&= \frac{1}{2 g^2 \xi}
  (\partial_\mu A_\mu^a)^2\,,\\ \mathcal{L}_\mathrm{ghost} &=
  \frac{1}{g^2}\partial_\mu \bar{c}^a D_\mu^{ab} c^b\,,
  \label{eq:lags}
  \end{split}
\end{equation}
respectively, where
\begin{equation}
  \begin{split}
    F_{\mu\nu}^a = \partial_\mu A^a_\nu - \partial_\nu A^a_\mu + f^{abc}A_\mu^bA_\nu^c
  \end{split}
\end{equation}
is the regular field strength tensor and
\begin{equation}
  D_\mu^\mathrm{F} = \partial_\mu + A_\mu^a T^a \,,\qquad D_\mu^{ab} = \delta^{ab} \partial_\mu - f^{abc} A_\mu^c
\end{equation}
are the regular covariant derivatives in the fundamental and adjoint representation, respectively, $g$
is the gauge coupling, $\xi$ the \qcd\ gauge parameter, and $\NF$ the
number of different quark flavors.  The flow equations are incorporated
by introducing Lagrange multiplier fields
\begin{equation}
  \begin{split}
    L_\mu^a(t,x)\qquad\text{and}\qquad
    \lambda_f(t,x)\,,\ \bar\lambda_f(t,x)\,,
  \end{split}
\end{equation}
of mass dimensions 3 and 5/2 that otherwise carry the same quantum
numbers as the flowed gluon and quark/antiquark
fields $B_\mu^a$ and $\chi,\bar\chi$, respectively.
Their Euler-Lagrange equations derived from
\begin{equation}
  \begin{split}
  \mathcal{L}_{B} &= -2 \int_0^\infty \mathrm{d}t \, \textrm{Tr} \left[
    L_\mu^a T^a \left(\partial_t B_\mu^b T^b -\mathcal{D}_\nu^{bc}
    G_{\nu \mu}^c T^b - \kappa \mathcal{D}_\mu^{bc} \partial_\nu B_\nu^c
    T^b \right) \right]\,,\\ \mathcal{L}_{\chi} & =
  \sum_{f=1}^{\NF}\int_0^\infty \mathrm{d}t \, \Big( \bar{\lambda}_f
  \left(\partial_t - \Delta + \kappa \left(\partial_\mu B_\mu^a\right)
  T^a \right) \chi_f \\&\hspace*{10em}
  + \bar{\chi}_f \left(\overleftarrow{\partial_t} -
  \overleftarrow{\Delta} - \kappa \left(\partial_\mu B_\mu^a\right) T^a
  \right) \lambda_f \Big)\,,
  \label{eq:Lagrangian_flow}
  \end{split}
\end{equation}
indeed lead to \eqn{eq:flow}\,\cite{Luscher:2011bx}.

Note that the Lagrangian \noeqn{eq:Lagrangian_complete} does not include
flowed ghost fields $d^a(t,x)$ and $\bar{d}^a(t,x)$. They arise in the
same way as the usual Faddeev-Popov ghosts $c^a(x)$ and $\bar{c}^a(x)$
due to gauge-fixing, and obey the initial condition
\begin{align}
  \left. d^a (t,x) \right|_{t=0} = c^a(x).
\end{align}
Similar to the ghosts of the regular gauge fields, they always form
closed loops as long as there are no external $d^a$ or $\bar{d}^a$
fields (which can be avoided by considering only physical degrees of
freedom in amplitudes with external gluons). As becomes clear later,
closed loops of only flowed fields vanish, so that one can omit $d^a$
and $\bar{d}^a$ already at the level of the Lagrangian.

\subsection{Perturbative solution of the flow equations}

Let us introduce the short-hand notation
\begin{equation}
  \begin{split}
    \int_p \equiv \int\frac{\dd^D p}{(2\pi)^D}\,,\qquad
    \int_x \equiv \int\dd^D x\,,
  \end{split}
\end{equation}
where it should be clear from the context whether an integration
variable is in position ($x,y,z,\ldots$) or momentum space
($p,k,q,\ldots$).

It is helpful to separate the flow equation \noeqn{eq:flow} of the flowed gauge
field $B_\mu^a(t,x)$ into a linear and a non-linear part:
\begin{equation}
  \begin{split}
  &\partial_t B_\mu^a = \partial_\nu \partial_\nu B_\mu^a + (\kappa - 1)
  \partial_\mu \partial_\nu B_\nu^a + R_\mu^a, \\ &R_\mu^a = 2 f^{abc}
  B_\nu^b \partial_\nu B_\mu^c - f^{abc} B_\nu^b \partial_\mu B_\nu^c +
  (\kappa - 1) f^{abc} B_\mu^b \partial_\nu B_\nu^c + f^{abe} f^{cde}
  B_\nu^b B_\nu^c B_\mu^d.
  \label{eq:flowsep}
  \end{split}
\end{equation}
The linear equation can be solved by introducing the integration kernel
\begin{equation}
  K_{\mu \nu} (t,x) = \int_p \frac{e^{\mathrm{i}px}}{p^2} \left( \left(
  \delta_{\mu \nu} p^2 - p_\mu p_\nu \right) e^{-tp^2} + p_\mu p_\nu \,
  e^{-\kappa tp^2} \right) \equiv \int_p e^{\ii px}\widetilde K_{\mu\nu}(t,p)\,,
  \label{eq:kmunu}
\end{equation}
which fulfills
\begin{equation}
  \lim_{t \to 0} K_{\mu\nu}(t,x) = \delta_{\mu\nu} \delta^{(D)}(x).
\end{equation}
Taking into account the initial condition \noeqn{eq:bound}, the full
solution of the flow equation is then given by
\begin{equation}
  B_\mu^a (t,x) = \int_y K_{\mu \nu}(t,x-y) A_{\nu}^a(y)+ \int_y
  \int_0^t \mathrm{d}s \, K_{\mu \nu}(t-s,x-y) R_\nu^a(s,y)\,,
\end{equation}
or, in momentum space,
\begin{equation}
  \widetilde{B}_\mu^a(t,p)= \int_x e^{-\mathrm{i}px} B_\mu^a(t,x)=
  \widetilde{K}_{\mu \nu}(t,p) \widetilde{A}_\nu^a(p)+ \int_0^t
  \mathrm{d}s \, \widetilde{K}_{\mu \nu}(t-s,p)
  \widetilde{R}_\nu^a(s,p).
  \label{eq:gluon_solution_momentum}
\end{equation}
By inserting the solution iteratively into itself, one can express the
Fourier transform of the non-linear part of \eqn{eq:flowsep} as
\begin{equation}
  \begin{split}
  \widetilde{R}_\mu^a(t,p) &=
  \int_{q,l,k} (2 \pi)^D \, \delta^{(D)}(p-q-l-k)\bigg[
   \delta^{(D)}(k)\cdot X_{2,\mu\nu\rho}^{abc}(q,l)\, \widetilde{B}_{\nu}^{b}(t,q)
      \widetilde{B}_{\rho}^{c}(t,l)\\&
  +
   X_{3,\mu\nu\rho\sigma}^{abcd}\, \widetilde{B}_{\nu}^{b}(t,q)
      \widetilde{B}_{\rho}^{c}(t,l)
      \widetilde{B}_{\sigma}^{d}(t,k)\bigg]\,,
  \end{split}
\end{equation}
where
\begin{equation}
  \begin{split}
  X_{2,\mu \nu \rho}^{abc}(q,l) &= -\mathrm{i} f^{a b c} \big[
  (l-q)_\mu \delta_{\nu \rho} + 2q{}_{\rho} \delta_{\mu \nu}- 2l{}_{\nu}
  \delta_{\mu \rho} + (\kappa - 1) ( q{}_{\nu} \delta_{\mu
    \rho} -l{}_{\rho} \delta_{\mu \nu}) \big]\,,\\
  X_{3,\mu \nu \rho \sigma}^{abcd} &= f^{abe}
  f^{cde}(\delta_{\mu \sigma} \delta_{\nu \rho} - \delta_{\mu \rho}
  \delta_{\nu \sigma}) + f^{ade} f^{bce}(\delta_{\mu \rho}
  \delta_{\nu \sigma}- \delta_{\mu \nu} \delta_{\rho \sigma})\\&\quad
  + f^{ace} f^{dbe}(\delta_{\mu \nu} \delta_{\rho \sigma}-
  \delta_{\mu \sigma} \delta_{\nu \rho}).
  \end{split}
\end{equation}
The structure of $X_3$ is identical to the four-gluon vertex of regular
\qcd. When formulating the Feynman rules later, $X_2$ and $X_3$ describe
the three- and four-point vertices of the flowed gluon fields.

The flow equation \noeqn{eq:flow} for the flowed quark fields can be
solved by again splitting it into a linear and a non-linear part,
\begin{align}
  \partial_t \chi =\partial_\mu \partial_\mu \chi + \Delta^\prime \chi
  \quad \textrm{with} \quad \Delta^\prime = (1-\kappa) \partial_\mu
  B_\mu^a T^a + 2 B_\mu^a T^a \partial_\mu + B_\mu^a B_\mu^b T^a T^b\,.
\end{align}
The linear equation is solved by the integration kernel
\begin{equation}
  K(t,x)=\int_p e^{\mathrm{i}px} e^{-tp^2} \equiv \int_p e^{\ii
    px}\widetilde K(t,p)\,,
  \label{eq:k}
\end{equation}
with the help of which we can write the full solution as
\begin{equation}
  \chi(t,x)= \int_y K(t,x-y)\psi(y) + \int_y \int_0^t \mathrm{d}s \, K(t-s,x-y) \Delta^\prime \chi(s,y).
\end{equation}
Here and in what follows, we suppress the flavor index $f$ unless
required for clarity.  The non-linear part of the Fourier-transformed
field
\begin{equation}
  \widetilde{\chi}(t,p)= \widetilde{K}(t,p)\widetilde{\psi}(p) + \int_0^t \mathrm{d}s \, \widetilde{K}(t-s,p)\widetilde{\Delta^\prime \chi}(s,p)
  \label{eq:quark_solution_momentum}
\end{equation}
can be expressed as
\begin{equation}
  \begin{split}
    \widetilde{\Delta^\prime \chi}(t,p) =&
    \int_{q,r} (2\pi)^D \delta^{(D)}(p-q-l-r)\bigg[
      \delta^{(D)}(l)\cdot Y_{1,\nu}^{b}(p,q,r)
      \widetilde{B}_{\nu}^{b}(t,q)
      \\&+
    \frac{1}{2}
    Y_{2,\nu\rho}^{bc}(p,q,l,r)
    \widetilde{B}_{\nu}^{b}(t,q)
    \widetilde{B}_{\rho}^{c}(t,l)\bigg] \widetilde{\chi}(t,r)\,,
  \end{split}
\end{equation}
where
\begin{equation}
  \begin{split}
  Y_{1,\nu}^b(q,r) &= \mathrm{i} \big( 2 r_\nu + (1-\kappa) q_\nu \big)
  T^b\,,\qquad
  Y_{2,\nu\rho}^{bc} = \delta_{\nu\rho} \bigl\{ T^{b}, T^{c} \bigr\}\,.
  \end{split}
\end{equation}
These expressions lead to the three- and four-point vertices of
the flowed quark fields. For $\bar{\chi}$ one proceeds analogously.

\subsection{Feynman rules}
\label{ssec:Feynman_rules}

\subsubsection*{(Flowed) Propagators}

Plugging in the solution of the flowed gluon field in momentum space
\eqn{eq:gluon_solution_momentum} into the two-point function, one
finds\,\cite{Luscher:2011bx}
\begin{align}
  \left\langle \widetilde{B}_\mu^a(t,p) \widetilde{B}_\nu^b(s,q)
  \right\rangle \Big|_{\textrm{LO}} &= \widetilde{K}_{\mu \rho}(t,p)
  \widetilde{K}_{\nu \sigma}(s,q) \left\langle \widetilde{A}_\rho^a(p)
  \widetilde{A}_\sigma^b(q) \right\rangle \nonumber \\ &= (2\pi)^D
  \delta^{(D)}(p+q) \, g^2\,D_{\mu\nu}^{ab}(p,t+s,\xi,\kappa)\,,
\end{align}
where
\begin{equation}
  \begin{split}
    D_{\mu\nu}^{ab}(p,t,\xi,\kappa) =
  \delta^{ab} \frac{1}{p^2} \bigg( \Big(
  \delta_{\mu \nu} - \frac{p_\mu p_\nu}{p^2} \Big) e^{-tp^2} + \xi
  \frac{p_\mu p_\nu}{p^2} e^{-\kappa tp^2} \bigg)\,,
  \end{split}
\end{equation}
and we have used the result for the fundamental gluon propagator,
\begin{equation}
\left\langle \widetilde{A}_\mu^a(p) \widetilde{A}_\nu^b(q) \right\rangle
\Big|_{\text{LO}} = (2\pi)^D \delta^{(D)}(p+q) \, g^2
D_{\mu\nu}^{ab}(p,0,\xi,0)\,.
\end{equation}
The factor $g^2$ is taken into account in the corresponding
vertices further below. Since the flowed gluon propagator coincides with the
fundamental gluon propagator at $t+s=0$, we can express both of them by
the same Feynman rule:
\begin{align}
  & \raisebox{-0.5em}{\includegraphics{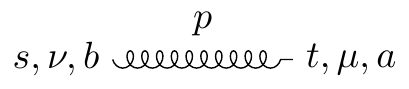}}
  & = &
  \begin{split}
    D_{\mu\nu}^{ab}(p,t+s,\xi,\kappa) \,.
  \end{split}
  & &
  \label{eq:flowedgluon}
\end{align}
We refer to this as the (flowed) gluon propagator. \eqn{eq:flowedgluon}
also applies to the mixed propagator $\langle \widetilde A \widetilde B
\rangle$, which is obtained by setting one of the two flow-time
variables to zero.

The same can be done for flowed quark fields. Inserting their momentum space
solution in \eqn{eq:quark_solution_momentum} into the two-point
function results in
\begin{align}
  \left. \left\langle \widetilde{\chi}^i_\alpha(t,p)
  \widetilde{\bar{\chi}}^j_\beta(s,q) \right\rangle
  \right|_{\text{LO}} &= \widetilde{K}(t,p) \widetilde{K}(s,q)
  \left\langle \widetilde{\psi}^i_\alpha(p)
  \widetilde{\bar{\psi}}^j_\beta(q) \right\rangle \nonumber \\ &=
  (2\pi)^D \delta^{(D)}(p+q)\, S^{ij}_{\text{F},\alpha\beta}(p,m,t+s)\,,
\end{align}
where
\begin{equation}
  \begin{split}
S^{ij}_\mathrm{F}(p,m,t) &= \delta^{ij} \frac{-i\slashed{p}+m}{p^2+m^2}\,
e^{-tp^2},
  \end{split}
\end{equation}
and we have used the result for the fundamental quark propagator
\begin{equation}
  \left\langle \widetilde{\psi}^i_\alpha(p)
  \widetilde{\bar{\psi}}^j_\beta(q) \right\rangle = (2\pi)^D
  \delta^{(D)}(p+q)\,S_{\text{F},\alpha\beta}^{ij}(p,m,0)\,.
\end{equation}
Since one can express the fundamental propagator through the flowed
propagator at vanishing flow times, both can be represented by the same
Feynman rule:
\begin{align}
  & \raisebox{-0.5em}{\includegraphics{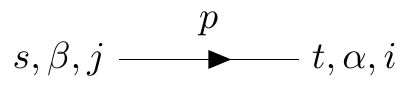}}
  & = &
  \begin{split}
    S_{\text{F},\alpha\beta}^{ij}(p,m,t+s) \,.
  \end{split}
  & &
\end{align}
We refer to this as the (flowed) quark propagator, which again
includes the mixed propagators
$\langle\psi\bar\chi\rangle$ and $\langle\chi\bar\psi\rangle$.

\subsubsection*{Flow lines}

Since there are no quadratic terms of the Lagrange multiplier fields
$L_\mu^a$, $\lambda$, and $\bar{\lambda}$ in \eqn{eq:Lagrangian_flow},
there are no propagators for these fields.  However, the
Lagrangian contains bilinear terms of a flowed field and a Lagrange
multiplier field.  Using standard methods, one derives the two-point
function at leading order as
\begin{equation}
  \left\langle B_\mu^a(t,x) L_\nu^b(s,y) \right\rangle = \delta^{ab}
  H_{\mu\nu}(t-s,x-y)\,,
\end{equation}
where $H_{\mu\nu}(t,x)$ obeys the equation
\begin{equation}
  \left[(\partial_t - \partial_\sigma \partial_\sigma) \delta_{\mu\nu} + (1 -
  \kappa) \partial_\mu \partial_\nu\right] H_{\nu\rho}(t,x) =
  \frac{1}{2\TR} \delta_{\mu\rho} \delta(t) \delta^{(D)}(x)\,,
\end{equation}
and $\TR$ is the usual color trace normalization.  Since the fundamental
gluon field $A_\mu^a(x) = B_\mu^a(0,x)$ does not couple to the Lagrange
multiplier field $L_\mu^a$, and all flow-time variables are positive,
one can impose the initial condition
\begin{equation}
  \begin{split}
  \left\langle B_\mu^a(t,x) L_\nu^b(s,y) \right\rangle
  \bigg|_{t=0} = 0 \quad \Rightarrow \quad
  H_{\mu\nu}(-s,x) = 0\,,
    \label{eq:bl}
  \end{split}
\end{equation}
so that the unique solution becomes
\begin{equation}
  H_{\mu\nu}(t,x) = \frac{1}{2\TR} \theta(t) K_{\mu\nu}(t,x)\,.
  \label{eq:hmunu}
\end{equation}
We refer to the $\langle BL\rangle$ bilinear as ``gluon flow line''.
The inverse of the factor $1/(2\TR)$ appears in the
corresponding ``flow vertices'' further below. We can therefore discard it
altogether and write
\begin{align}
  & \raisebox{-1.2em}{\includegraphics{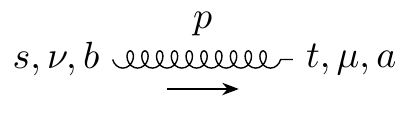}}
  & = &
  \begin{split}
    \delta^{ab} \, \theta(t-s)\,\widetilde K_{\mu\nu}(t-s,p) \,,
  \end{split}
  & &
  \label{eq:blfr}
\end{align}
where $\widetilde K_{\mu\nu}(t,p)$ has been defined in \eqn{eq:kmunu},
and the adjacent arrow indicates the direction towards increasing
flow time as implied by the $\theta$-distribution.  As opposed to an
actual propagator, the gluon flow line is a regular function for all
$p$.

Similarly, one determines the mixed fermionic two-point function
at leading order as
\begin{equation}
  \left\langle \chi_\alpha^i(t,x) \bar{\lambda}_\beta^j(s,y) \right\rangle =
  \delta_{\alpha\beta} \delta^{ij} G(t-s,x-y)\,,
\end{equation}
where $G(t,x)$ obeys
\begin{equation}
  (\partial_t - \partial_\mu \partial_\mu) G(t,x) = \delta(t) \delta^{(D)}(x)\,,
\end{equation}
with the condition
\begin{equation}
  \left\langle \chi_\alpha^i(t,x) \bar{\lambda}_\beta^j(s,y)
  \right\rangle \bigg|_{t=0} = 0 \quad \Rightarrow \quad G(-s,x)= 0\,.
\end{equation}
The unique solution reads
\begin{equation}
  G(t,x) = \theta(t) K(t,x)\,.
\end{equation}
The Fourier transformed expression
defines the ``fermion flow line'' Feynman rule, where the
$\theta$-distribution again imposes a direction as indicated by the adjacent arrow
pointing towards increasing flow time:
\begin{align}
  & \raisebox{-1.2em}{\includegraphics{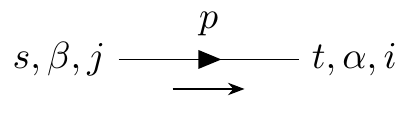}}
  & = &
  \begin{split}
    \delta_{\alpha\beta} \delta_{ij} \, \theta(t-s) \, \widetilde K(t-s,p) \,,
  \end{split}
  & &
  \label{eq:chibarlambda}
\end{align}
where $\widetilde K(t,p)$ has been defined in \eqn{eq:k}.  As usual, the
arrow \textit{on} the fermion line denotes the ``charge flow'' of the
fermion. In this way, we have a unified Feynman rule for both the
$\langle\chi\bar\lambda\rangle$ and the $\langle\lambda\bar\chi\rangle$
bilinear, where the latter can be obtained analogously as above
and simply corresponds to reversing the direction of the charge flow:
\begin{align}
  & \raisebox{-1.2em}{\includegraphics{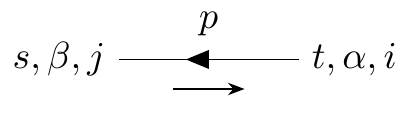}}
  & = &
  \begin{split}
    \delta_{\alpha\beta} \delta_{ij} \, \theta(t-s) \, \widetilde K(t-s,p) \,.
  \end{split}
  & &
  \label{eq:lambdabarchi}
\end{align}
Since $\widetilde K(t,p)$ only depends on $p$ only quadratically, the
momentum direction for the fermion flow lines is irrelevant.

\subsubsection*{Flow vertices}

Since $\mathcal{L}_B$ and $\mathcal{L}_\chi$ of
\eqn{eq:Lagrangian_flow} are proportional to a Lagrange multiplier
field, the resulting vertices always involve at least one flow
line. Such vertices are always associated with a flow-time
parameter which is integrated over. We denote them by ``flow vertices''
and represent them by empty circles in Feynman diagrams. The
corresponding Feynman rules can be derived straightforward.
In this paper, we define Feynman rules by assuming all momenta to be
outgoing.

\begin{figure}
  \begin{equation*}
    \begin{gathered}
      \includegraphics{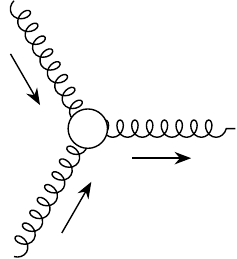}
    \end{gathered}
    \qquad
    \begin{gathered}
      \includegraphics{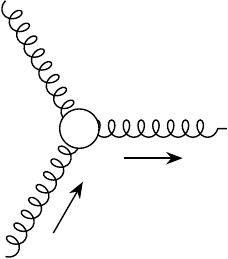}
    \end{gathered}
    \qquad
    \begin{gathered}
      \includegraphics{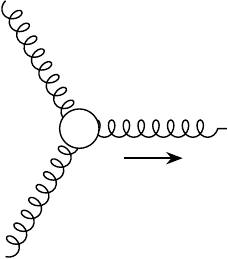}
    \end{gathered}
  \end{equation*}
  \caption{The three versions of the vertex $X_2$. Lines with an
    adjacent arrow denote flow lines, the others are flowed (or regular)
    gluons.}
  \label{fig:3gcomb}
\end{figure}

The three-point gluon flow vertex is governed by $X_2$:
\begin{align}
  &
  \begin{gathered}
    \includegraphics{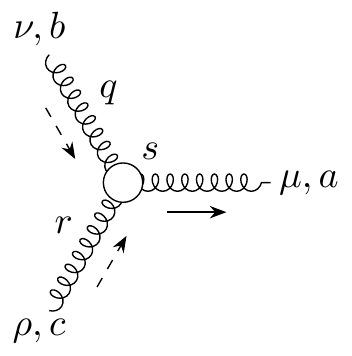}
  \end{gathered}
  & = &
  \begin{split}
    -\mathrm{i} g f^{abc} \int_0^\infty \mathrm{d}s \, \big( & \delta_{\nu\rho} (r-q)_\mu + 2 \delta_{\mu\nu} q_\rho - 2 \delta_{\mu\rho} r_\nu \\
    & + (\kappa-1) (\delta_{\mu\rho} q_\nu  - \delta_{\mu\nu} r_\rho) \big) \,. \\
  \end{split}
  & &
  \label{eq:3gvert}
\end{align}
The interaction terms of $\mathcal{L}_B$ involve exactly one Lagrange
multiplier field $L(s)$, cf.\ \eqn{eq:Lagrangian_flow}.  According to
\eqs{eq:bl} and \noeqn{eq:blfr}, it must be contracted with a flowed
gluon field $B(t)$ at larger flow time $t>s$. Thus, the vertex in
\eqn{eq:3gvert} contains exactly one outgoing flow line. On the other
hand, each of the two flowed gluon fields $B(s)$ in the interaction
terms of $\mathcal{L}_B$ can be contracted either with a Lagrange
multiplier field $L(t')$ at \textit{smaller} flow time $t'<s$, resulting
in an ingoing flow line, or with another flowed gluon field $B(s')$. In
this sense, the Feynman rule \noeqn{eq:3gvert} actually represents three
vertices, displayed in \fig{fig:3gcomb}: They all contain one outgoing
flow line, while each of the other two lines can either be an ingoing
flow line or a flowed gluon propagator.\footnote{We focus on the
  calculation of Green's functions in this paper; if one calculates
  amplitudes, the lines could also represent external ``particles'', of
  course.}  The dashed arrows in \eqn{eq:3gvert} indicate lines which
can be both a flow line or a propagator, while the solid arrows always
denote flow lines.

Similarly, the four-point gluon flow vertex is governed by $X_3$:
\begin{align}
  &
  \begin{gathered}
    \includegraphics{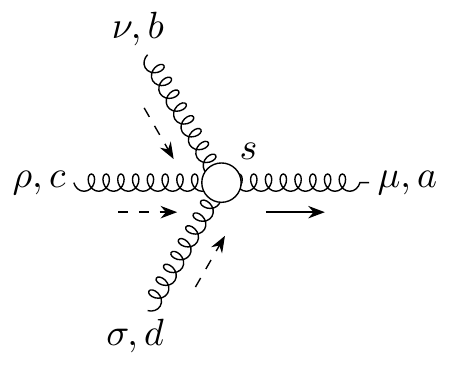}
  \end{gathered}
  & = &
  \begin{split}
    -g^2 \int_0^\infty \mathrm{d}s \, \big( & f^{abe} f^{cde} (\delta_{\mu\rho}\delta_{\nu\sigma} - \delta_{\mu\sigma}\delta_{\nu\rho} ) \\
    & + f^{ace} f^{bde} (\delta_{\mu\nu}\delta_{\rho\sigma} - \delta_{\mu\sigma}\delta_{\nu\rho} ) \\
    & + f^{ade} f^{bce} (\delta_{\mu\nu}\delta_{\rho\sigma} - \delta_{\mu\rho}\delta_{\nu\sigma} ) \big) \,.
  \end{split}
  & &
  \label{eq:4gvert}
\end{align}
Again, there is exactly one outgoing flow line, while the other three
lines are either flowed gluons or incoming flow lines. This means that
the Feynman rule \noeqn{eq:4gvert} actually represents four different
vertices.

Note that the factor $2\TR$, arising from the trace in
\eqn{eq:Lagrangian_flow}, has been discarded in both vertices in accordance with the
normalization of \eqn{eq:blfr}.

Similar considerations applied to $\mathcal{L}_\chi$ lead to vertices
involving flowed quark fields. For example, the quark flow vertex with
one gluon is described by the following Feynman rule:
\begin{align}
  &
  \begin{gathered}
    \includegraphics{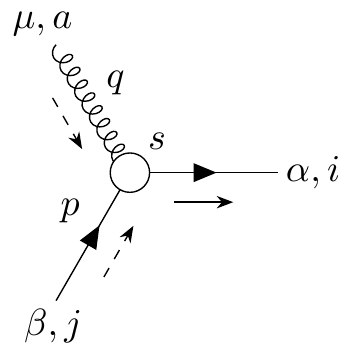}
  \end{gathered}
  & = &
  \begin{split}
    \mathrm{i} g \, \delta_{\alpha\beta} \big( T^a \big)_{ij} \int_0^\infty \mathrm{d} s \, \big( 2 p_\mu + (1-\kappa) q_\mu \big) \,.
  \end{split}
  & &
\end{align}
In this case, the fermion line with outgoing charge flow is also
outgoing in the gradient flow, while the other two lines can be either
flowed propagators, or ingoing flow lines. Thus, this Feynman rule
actually represents four different vertices.

A complete list of the Feynman rules can be found in
Appendix\,\ref{sec:frules}.

Let us summarize the Feynman rules for the \gff\ and point out a few
more features:
\begin{enumerate}
  \item Propagators and flow lines always carry an exponential factor
    $\mathrm{e}^{\pm tp^2}$ for each vertex, where $t$ is the flow time
    of the vertex.
  \item Flow lines always start at a flow vertex.  They end either at
    another flow vertex or an external operator. The $\theta$-distribution
    $\theta (t-s)$ implies a flow-time direction, which we denote by an
    adjacent arrow in Feynman diagrams.
  \item A flow vertex always has an outgoing flow line connected to it.
    The other lines are either incoming flow lines, or flowed
    propagators.  Each flow vertex is defined at a
    flow time $s$ and implies an integration $\int_0^\infty
    \mathrm{d}s$.  The $\theta$-distribution of the outgoing flow line
    restricts the integration to a finite upper limit $t$,
    i.e.\ $\int_0^t \mathrm{d}s$.
  \item Diagrams with closed flow-line loops vanish, because the
    integration interval shrinks to a point. Hence, only diagrams whose
    flow lines form trees contribute to any observable.
\end{enumerate}

\section{Automated implementation}
\label{sec:implementation}
\subsection{Generation of Feynman diagram expressions}

We generate the Feynman diagrams including symmetry factors and signs
(from closed fermion loops) with the help of the program
\texttt{qgraf}\,\cite{Nogueira:1991ex,Nogueira:2006pq}. In the notation
of this program, the propagators for the gluon, the ghost, and a single
quark flavor can be defined as
\begin{verbatim}
[g,g,+], [c,C,-], [fq,fQ,-].
\end{verbatim}
The sign in the third entry of each square bracket
denotes whether the particle is a boson or a fermion.

The flow lines for the gluon and the quark are implemented by
introducing separate fields \texttt{b}, \texttt{fr}, and \texttt{fs} for
$L$, $\lambda$, and $\bar\lambda$, respectively. They are implemented as
\begin{verbatim}
[b,B,+], [fr,fR,-], [fs,fS,-],
\end{verbatim}
where the latter two represent the $\langle\chi\bar\lambda\rangle$ and
$\langle\lambda\bar\chi\rangle$ bilinears (see \eqs{eq:chibarlambda} and
\noeqn{eq:lambdabarchi}). With these fields, we can then define the
regular as well as the flow vertices. For example, for the trilinear
flow vertex of the pure gauge theory defined in \eqn{eq:3gvert}, we
define
\begin{verbatim}
[B,b,b], [B,b,g], [B,g,g],
\end{verbatim}
which corresponds to the three combinations of
\fig{fig:3gcomb}.

Already in regular \qcd\ it is convenient to separate the color
structure from the rest of the calculation. This becomes non-trivial in
cases which involve the four-gluon vertex. It is thus convenient to
introduce an auxiliary ``particle'' $\Sigma_{\mu\nu}^{a}$ whose
``propagator'' is given by\,\cite{Chetyrkin-priv}
\begin{align}
  &
  \begin{gathered}
    \includegraphics{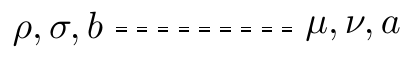}
  \end{gathered}
  & = &
  \begin{split}
    \delta^{ab}\delta_{\mu\rho}\delta_{\nu\sigma} \,.
  \end{split}
  & &
\end{align}
The four-gluon vertex can then be replaced by a trilinear $gg\Sigma$
vertex whose Feynman rule reads
\begin{align}
  &
  \begin{gathered}
    \includegraphics{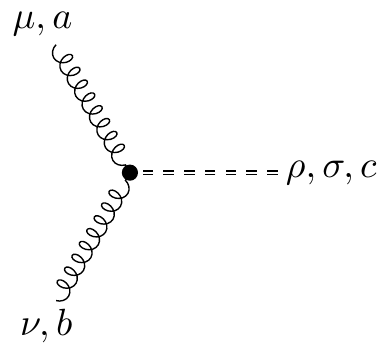}
  \end{gathered}
  & = &
  \begin{split}
    \frac{\mathrm{i}g}{\sqrt{2}} f^{abc}\left(\delta_{\mu\rho}\delta_{\nu\sigma}-\delta_{\nu\rho}\delta_{\mu\sigma}\right) \,.
  \end{split}
  & &
\end{align}
This allows to factorize the color factor off all Feynman diagrams,
albeit at the cost of increasing their number. We proceed
correspondingly for the quartic gluon flow vertices by introducing an
additional $\Sigma$ particle which is directed in flow time. It forms
trilinear vertices with gluons and gluon flow lines, as shown
explicitly in Appendix\,\ref{sec:frules}. Again, they contain
exactly one outgoing ($\Sigma$ or gluon) flow line.

By default, \texttt{qgraf} also generates diagrams with closed flow-line
loops.  We use a \texttt{Perl} script to parse the \texttt{qgraf} output
in order to eliminate such diagrams before the actual calculation. They
vanish trivially algebraically though, as discussed earlier.  In
the same way we multiply diagrams by a factor $\NF$ for each closed
fermion loop.  We then process the diagrams with the help of
\texttt{q2e/exp}\,\cite{Harlander:1997zb, Seidensticker:1999bb} in order
to insert the Feynman rules and convert the diagrams to \texttt{FORM}
code.

Within \texttt{FORM}\,\cite{Vermaseren:2000nd, Kuipers:2012rf}, we
contract the Lorentz indices, take the fermion traces, and simplify the
resulting expressions to a standard form of scalar integrals, as defined below.
  The color factor is evaluated separately with the
help of the \texttt{color} package\,\cite{vanRitbergen:1998pn}.

For example at the three-loop level (as needed for the results in this paper)
the flow-time integrals can be described in the following form
\begin{equation}
\begin{split}
&I(t,\mathbf{n},\mathbf{a},\mathbf{c},D) =
  \left(\prod_{r=1}^N\int_0^{t_{r}^\text{up}}\dd
  t_r\,t_r^{c_r}\right)\int_{p_1,p_2,p_3} \frac{\exp[\sum_{k,i,j}
      a_{kij} t_k p_i \cdot p_j] }{p_1^{2n_1}p_2^{2n_2}p_3^{2n_3}
    p_4^{2n_4}p_5^{2n_5}p_6^{2n_6}}\,,
\label{eq:intrep}
\end{split}
\end{equation}
where
\begin{equation}
\begin{split}
\mathbf{n} &= \{n_1,\ldots,n_6\}\,,\qquad\mathbf{c} = \{c_1,\ldots,c_N\}\,,\\
\mathbf{a} &= \{a_{kij}:k=0,\ldots, N\,;\,i=1,2,3\,;j=1,2,3\}\,,\\
\end{split}
\end{equation}
are sets of integers, $N\leq 4$, $t_0\equiv t$, and the upper limits for
the flow-time integrations are linear combinations of the other
flow-time variables, $t_r^\text{up}=t_r^{\text{up}}(t_0,\ldots,
t_{r-1})$. Initially, all the $c_i$ are zero. It is helpful to introduce
these parameters for the subsequent discussion though.  The momenta
$p_4$, $p_5$, $p_6$ are linear combinations of the integration momenta
$p_1$, $p_2$, $p_3$.  For the three-loop quantities considered
in this paper, the flow time $t$ is the only dimensionful external scale.
A generalization of this notation to a different number of loops and to additional external
scales is straightforward.

\subsection{\abbrev{IBP} reduction of flow-time loop integrals}\label{sec:reduction}

In the first step, we reduce all occurring integrals to so-called master
integrals by employing integration-by-parts (\abbrev{IBP}) identities\,\cite{Tkachov:1981wb,Chetyrkin:1981qh}.
They are based on the
observation that integrals over total derivatives vanish in dimensional
regularization. Applying the operator
\begin{equation}
  \begin{split}
    \mathcal{D}_{ij} \equiv \frac{\partial}{\partial p_i}\cdot p_j =
    \delta_{ij}\,D+ p_j\cdot \frac{\partial}{\partial p_i}
    \label{eq:dij}
  \end{split}
\end{equation}
to the integrand of $I(t,\mathbf{n},\mathbf{a},\mathbf{c},D)$ in
\eqn{eq:intrep} thus results in a vanishing integral for all
$i,j\in\{1,2,3\}$. On the other hand, explicitly acting with the
derivative on the integrand leads to a sum of integrals where some of the
indices $n_i$ and $c_r$ are shifted by $\pm 1$.\footnote{To be
  precise: At most one $n_i$ is shifted by $-1$ and at most one of the
  $n_i$ and at most one of the $c_r$ are shifted by $+1$.} Repeated
application of the $\mathcal{D}_{ij}$ thus leads to linear relations
among integrals with different $\mathbf{n}$ and $\mathbf{c}$.
However, the number of flow-time integrations and the exponential
function in the integrand still remain unaltered by this procedure.

We may apply an analogous strategy for the flow-time parameters though.
Inserting a derivative with respect to one of the flow-time integration
variables, one arrives at the sum of two integrals with fewer flow-time
integrations and an altered exponential,
\begin{equation}
  \int_0^{t_{r}^\text{up}}\dd t_r \deriv{}{t_r} f(t_r,\ldots) = f(t_{r}^\text{up},\ldots)-f(0,\ldots)\,.
  \label{eq:ibpflow}
\end{equation}
On the other hand, explicit evaluation of the derivative at the
integrand level either reduces one of the indices $c_k$ or one of the indices $n_i$
by one. One therefore arrives at linear relations among integrals with
different $\mathbf{n}$, $\mathbf{c}$, $\mathbf{a}$, and different
number of flow-time integrations $N$.

Applying the above operations to ``seed integrals'', i.e., integrals
with fixed numerical values for the $\mathbf{n}$, $\mathbf{c}$, and
$\mathbf{a}$, allows one to build a system of linear relations among the
$I(t,\mathbf{n},\mathbf{a},\mathbf{c},D)$. Defining an ordering
(complexity) criterion allows one to solve the system using a Gaussian elimination
type reduction for a minimal set of integrals (master integrals) \cite{Laporta:2001dd}.
Through this procedure all integrals of the type defined in \eqn{eq:intrep}
are reduced to the minimal set of master integrals, which are simplest
by the ordering criterion introduced. The system of linear relations
is process dependent and for the observables in this study we construct it
in \texttt{Mathematica}\,\cite{Mathematica11.3}. We use the specialized software
\texttt{Kira}\,\cite{Maierhoefer:2017hyi,Maierhofer:2018gpa} for the reduction of
such linear equation systems to solve it.

While \texttt{Kira} alone is sufficient
for the reduction of all our integrals at the two-loop level (see
e.g.\ \citere{Harlander:2018zpi}), we find that the algebraic solution
of the system at the three-loop level would require more than
750\,GiB of \abbrev{RAM} and thus exceeds our available computing
resources.\footnote{This calculation was performed before the release of
  \texttt{Kira 1.2}\,\cite{Maierhofer:2018gpa}. It is well possible that
  these statements could change with the newly implemented features.}
Using finite field and
reconstruction techniques, which
over the last decade have gained an increased use for higher-order
perturbative calculations (see,
e.g.\ \citeres{Kauers:2008zz,Kant:2013vta,Peraro:2016wsq}), one can improve on the required computational resources as
follows.

Since $t$
is the only dimensionful scale and can thus be factored out, the
coefficients of the integrals in the linear system are simply
polynomials in $D$. The individual steps for the reduction of the linear system only involve
elementary arithmetic operations, which means that the coefficients of the master
integrals are rational functions in $D$. One can then solve the linear system
numerically over a finite field using a sufficiently large number of different integer
values for $D$ (``probes'') and reconstruct the rational functions in $D$ exactly. With this approach
the size of the coefficients remain simple numbers and the requirements on computational
resources can be improved.

While \texttt{Kira} already uses \texttt{pyRed} as a first step to remove
linearly dependent equations over a finite field, in our approach \texttt{pyRed} is used
to reduce the system itself multiple times over a finite field.
The resulting numbers are processed with the
library \texttt{FireFly}\,\cite{Klappert:2019emp}, which provides an
efficient implementation of interpolation\,\cite{Zippel:1990,Cuyt:2011}
and rational-reconstruction
algorithms\,\cite{Wang:1981,Monagan:2004}.\footnote{We remark that
  \texttt{FireFly} also works for multi-variate rational functions.}  In
our case, we require 201 probes, chosen from three different finite
fields, defined as prime fields $\mathbb{Z}_p$ with $p$ a 63-bit prime
number, to reconstruct all coefficients, plus one additional probe in a
fourth prime field to verify the reconstruction. For each probe, the
solution of the system with \texttt{Kira} now just requires about 70\,GiB of
\abbrev{RAM} and takes about three \abbrev{CPU} hours.  In total, the reduction took
less than three days, using ten threads on two Intel Xeon Gold 6138
processors.

For the observables considered further below in this paper at the three-loop level we start
with a total of 3195 integrals and can reduce them to a minimal set of 188 master integrals.\footnote{When our
	observables are expressed in terms of the master integrals, the dependence of six of them drops out.}
Their numerical evaluation is described in the next section.

\subsection{Numerical computation of flow-time loop integrals}

As a first step of the numerical evaluation of the gradient-flow integrals given by \eqn{eq:intrep},
we express the propagators through Schwinger-parameter integrals and map them from $x\in(0,\infty)$ to
$y\in(0,1)$ using the simple transformation $x=y/(1-y)$.
Momentum integrations are performed as $D$-dimensional Gaussian integrals after a diagonalization.
Similarly, all
flow-time integrations are mapped to the unit interval with simple
linear transformations. The overall result is an integral over a unit hypercube.

The integrand typically involves a number of (overlapping)
singularities, which we factorize using
\texttt{FIESTA}\,\cite{Smirnov:2013eza}, an implementation of the sector
decomposition algorithm\,\cite{Binoth:2003ak}.  As opposed to regular
Feynman integrals where sector decomposition is typically employed in
combination with Feynman parameterization, it appears that we cannot
restrict the singularities to the lower integration bound
only.\footnote{To identify cases where singularities appear at the upper
  bound, we determine the degree of divergence for each subset of
  integration variables at the level of the Schwinger parameterization
  \cite{Panzer:2014gra}.} If an integral involves singularities both at the
lower and the upper bound, we split all integration intervals in the
middle, and map the singularities to the lower bound by an appropriate
change of variable.

The sector decomposed integrals obtained from \texttt{FIESTA} are then
integrated with our implementation of fully symmetric integration rules
of order 13\,\cite{GenzMalik,Harlander:2016vzb}, which can handle
integrable logarithmic-like singularities at the integration boundaries
very well. We perform all arithmetic with 256-bit precision using the
\texttt{MPFR} library\,\cite{Fousse:2007}, and used a local adaptive
bisection in the direction of the largest fourth difference. A high
precision arithmetic turns out to be necessary to achieve a relative
numerical accuracy of $10^{-10}$ or better. The integration uncertainty
is estimated by the difference between the integration results of rules
of order 13 and 11.

As a check of our results further below, we apply the numerical integration method
to our unreduced set of $3195$ integrals as well as to the set of $188$ master integrals.
Finding full agreement for the results obtained with both sets within numerical uncertainties
serves as a check of the numerical integration as well as for the reduction procedure.

\paragraph{Analytical computation.}
For some master integrals we can find analytical results either
through elementary methods with the help of
\texttt{Mathematica}\,\cite{Mathematica11.3} or by using
\texttt{HyperInt} \cite{Panzer:2014caa}.
The integration with \texttt{Mathematica} sometimes yields hypergeometric functions which can be expanded with the package \texttt{HypExp}\,\cite{Huber:2005yg,Huber:2007dx}.
For example as shown further below, we
find analytical results for all contributions which contain at least one factor
of $\TR$ in the color structure. A fully
analytical calculation of integrals with other color factors requires a more
detailed investigation. However, for all practical purposes, the
numerical results provided in this paper are (more than) sufficient.

\section{Observables}
\label{sec:observables}

Among the simplest quantities one can consider within the \gff\ are
vacuum expectation values of gauge-invariant operators at finite flow
time. It is one of the remarkable properties of the \gff\ that these
operators do not require any renormalization beyond that of regular
\qcd, and that of the involved flowed fields. For the gauge coupling and
the quark masses, the \msbar\ renormalization is given by the
replacement
\begin{equation}
  \begin{split}
    g\to g_0\equiv \left(\frac{\mu\mathrm{e}^{\gamma_\mathrm{E}/2}}{\sqrt{4\pi}}\right)^\ep
    Z_g(\alphas(\mu))\,g(\mu)\,,\qquad m_f\to m_{f,0}\equiv
    Z_m(\alphas(\mu))\,m_f(\mu)
    \label{eq:renorm}
  \end{split}
\end{equation}
in \eqn{eq:lags}, where $\mu$ is the renormalization scale,
$\alphas=g^2/(4\pi)$, and $\EulerGamma=0.5772\ldots$ the
Euler-Mascheroni constant.  Through the perturbative order required in
this paper, the renormalization constants are given by
\begin{equation}
  \begin{split}
    Z_g(\alphas) &=
    1 -
    \frac{\alphas}{4\pi}\frac{\beta_0}{2\ep} +
    \left(\frac{\alphas}{4\pi}\right)^2\left(\frac{3\beta_0^2}{8\ep^2}
    -\frac{\beta_1}{4\ep}\right) + \order{\alphas^3}\,, \\ Z_m(\alphas) &=
    1 -\frac{\alphas}{4\pi}\frac{\gamma_{m,0}}{2\ep}
    +\left(\frac{\alphas}{4\pi}\right)^2\left[\frac{1}{\ep^2}\left(
      \frac{\gamma_{m,0}^2}{8} + \frac{\beta_0\gamma_{m,0}}{4}\right)
      -\frac{\gamma_{m,1}}{4\ep}\right] + \order{\alphas^3}\,,
    \label{eq:rgconsts}
  \end{split}
\end{equation}
with
\begin{equation}
\begin{aligned}
      \beta_0 &= \frac{11}{3} \CA - \frac{4}{3} \TF\,, &
      \beta_1 &= \frac{34}{3} \CA^2 - \left( 4 \CF + \frac{20}{3} \CA
      \right) \TF \,,\\
      \gamma_{m,0} &= 6 \CF\,, &
      \gamma_{m,1} &= \frac{97}{3} \CA \CF + 3 \CF^2
      - \frac{20}{3} \CF \TF \,.
    \label{eq:anombg}
\end{aligned}
\end{equation}
$\CF$ and $\CA$ are the quadratic Casimir eigenvalues of the fundamental
and the adjoint representation of the gauge group,
respectively. Furthermore, $\TF=\TR \NF$, with $\NF$ the number of quark
flavors, and $\TR$ the trace normalization in the fundamental
representation. For SU($\NC$), it is $\CF=(\NC^2-1)/(2\NC)$, $\CA=\NC$,
and $\TR=1/2$.

The flowed gauge field $B^a_\mu(t,x)$ does not require renormalization,
so that for example matrix elements of the gluon action density,
\begin{align}
  E (t,x) \equiv \frac{1}{4} G_{\mu \nu}^a(t,x) G_{\mu \nu}^a(t,x)\,,
  \label{eq:def_E}
\end{align}
are finite after just the renormalization of $g$ and $m_f$.
This allows for a direct comparison of results
obtained in different regularization schemes (lattice and perturbation
theory, for example).

On the contrary, flowed quark fields require a renormalization factor
$Z_\chi^{1/2}(\alphas)$ in order to render Green's functions finite. In
the \msbar\ scheme, it
is\,\cite{Luscher:2010iy,Harlander:2018zpi}
\begin{equation}
  \begin{split}
        Z^{-1}_\chi(\alphas) &= 1
        -\frac{\alphas}{4\pi}\frac{\gamma_{\chi,0}}{2\ep}
        +\left(\frac{\alphas}{4\pi}\right)^2\left[\frac{1}{\ep^2}\left(
          \frac{\gamma_{\chi,0}^2}{8} +
          \frac{\beta_0\gamma_{\chi,0}}{4}\right)
          -\frac{\gamma_{\chi,1}}{4\ep}\right] + \order{\alphas^3}\,,
  \end{split}
\end{equation}
with
\begin{equation}
  \begin{split}
  \gamma_{\chi,0} &= 6\,\CF\,,\\
  \gamma_{\chi, 1} &=
  \left( \frac{223}{3} - 16 \ln 2 \right) \CA \CF - \left( 3 + 16
  \ln 2 \right) \CF^2 - \frac{44}{3} \CF \TF\,.
  \end{split}
\end{equation}

The quantity
\begin{align}
  S (t,x) \equiv Z_\chi\,\sum_{f=1}^{\NF}\bar{\chi}_f(t,x) \chi_f(t,x)
  \label{eq:def_S}
\end{align}
thus acquires an anomalous dimension, which prevents a direct comparison
of results from different regularization schemes. Alternatively, one may
work with ``ringed quark fields''\,\cite{Makino:2014taa}, which amounts to using
\begin{equation}
  \begin{split}
    \mathring{Z}_\chi(t,\mu) = -\frac{2\NC \NF}{(4\pi t)^2}\, \langle
    R(t)\rangle^{-1}\,,\quad\mbox{with}\quad R(t,x) =
    \sum_{f=1}^{\NF}\bar{\chi}_f(t,x) \overleftrightarrow{\slashed{\mathcal{D}}}^\mathrm{F} \chi_f(t,x)
  \label{eq:Zring}
  \end{split}
\end{equation}
instead of $Z_\chi$ in order to renormalize the quark fields, where
\begin{equation}
  \overleftrightarrow{\mathcal{D}}^\mathrm{F} = \mathcal{D}^\mathrm{F} - \overleftarrow{\mathcal{D}}^\mathrm{F} \,.
\end{equation}
This
corresponds to a ``physical'' renormalization scheme, which means that
the anomalous dimension of the operator
\begin{equation}
  \begin{split}
    \mathring{S}(t,x) = \zeta_\chi(t,\mu)\,S(t,x)\,,\qquad\mbox{with}\quad
    \zeta_\chi(t,\mu) \equiv Z_\chi^{-1}\mathring{Z}_\chi(t,\mu)\,,
    \label{eq:zetachi}
  \end{split}
\end{equation}
vanishes. In \eqn{eq:Zring}, we used the notation
\begin{equation}
  \begin{split}
    \langle \mathcal{O}(t)\rangle \equiv \int\dd^4 x\,
    \langle \mathcal{O}(t,x)\rangle
  \end{split}
\end{equation}
for the zero-momentum Fourier transform of the vacuum expectation value
of an operator $\mathcal{O}(t,x)$, which we adopt in the following.

The Feynman rules for the operators $E(t,x)$, $S(t,x)$, and $R(t,x)$
introduced above are obtained in the same manner as those described in
\sct{ssec:Feynman_rules}. Expressing them in terms of the flowed fields
$B(t,x)$ and $\chi_f(t,x)$, the operator $E(t,x)$ results in a bilinear,
trilinear, and a quartic gluon vertex as displayed in
\eqs{fr:egg}--\noeqn{fr:egggg}, $S(t,x)$ corresponds to the single
bilinear quark vertex given in \eqn{fr:sqq}, and $R(t,x)$ to the
bilinear quark vertex of \eqn{fr:rqq} and the quark-gluon vertex of
\noeqn{fr:rqqg}. Similar to the flow vertices, the lines attached to
these vertices are depicted with dashed arrows, indicating that they can
be both flow lines and propagators.  In contrast to the flow vertices,
the flow time of all lines in these vertices is directed towards the
corresponding operator.  Therefore, these vertices set the largest flow
time in the Feynman diagrams.

\begin{figure}
  \begin{center}
    \begin{tabular}{ccc}
         \includegraphics[]{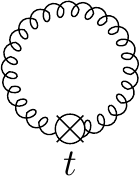} (a) &
         \includegraphics[]{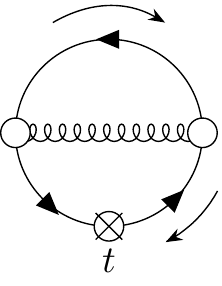} (b) &
         \includegraphics[]{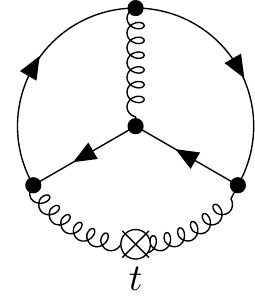} (c) \\[2em]
         \includegraphics[]{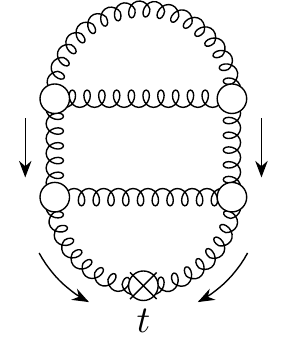}\hspace*{-1em}
         (d)\hspace*{1em} &
         \includegraphics[]{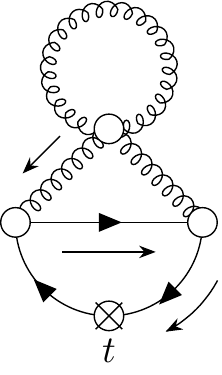} (e) &
         \includegraphics[]{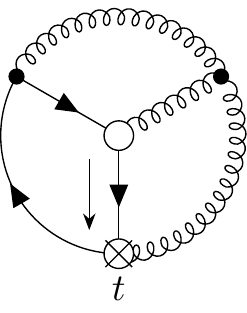} (f) \\
    \end{tabular}
  \end{center}
	\caption{The only diagram contributing to $\langle E(t)\rangle$
          at one loop (a); a \two-loop diagram for $\langle S(t)\rangle$
          and $\langle R(t)\rangle$ (b); and four three-loop diagrams:
          two diagrams for $\langle E(t)\rangle$ (c,d); one diagram for
          $\langle S(t)\rangle$ and $\langle R(t)\rangle$ (e); and one
          diagram for $\langle R(t)\rangle$ (f).}
	\label{fig:example_diagrams}
\end{figure}

Using the methods described in \scts{sec:pert} and \ref{sec:implementation}, we can calculate the
vacuum expectation values of $E(t,x)$, $S(t,x)$, and $R(t,x)$ through
three loops. Sample diagrams for these quantities are shown in
\fig{fig:example_diagrams}. The number of diagrams in each case is given
in Table\,\ref{tab:numdias}; recall, however, that this takes into
account the re-writing of the \four-gluon vertex into trilinear vertices
as described in \sct{sec:implementation}. Not included in this number
are diagrams with closed flow-line loops, but integrals with scale-less
sub-loops are counted in.

\begin{table}
  \begin{center}
    \caption[]{\label{tab:numdias} The number of Feynman diagrams
      contributing to the quantities computed in this section. At
      \three-loop level, the number for ``quenched \qcd'' (marked as
      $\TR=0$) is shown separately.}

    \begin{tabular}{|c||c|c|c|c|}
      \hline
      \#loops & 1 & 2 & 3 & 3  ($\TR=0$)\\\hline\hline
      $\langle E(t)\rangle$ & 1 & 11 & 232 & 211 \\\hline
      $\langle S(t)\rangle$ & 1 & 8 & 210 & 202 \\\hline
      $\langle R(t)\rangle$ & 1 & 11 & 311 & 300 \\\hline
    \end{tabular}
  \end{center}
\end{table}

\subsection{Results for the gluon condensate at three loops}

We write the perturbative result for the gluon condensate as
\begin{equation}
  \langle E(t) \rangle = \frac{3\alphas}{4\pi t^2} \frac{\NA}{8}
  \left[e_0 + \frac{\alphas}{4\pi} e_1 + \left(\frac{\alphas}{4\pi}\right)^2 e_2 +
    \mathcal{O}(\alphas^3)\right] + \mathcal{O}(m)\,,
  \label{eq:et}
\end{equation}
where $\alphas=\alphas(\mu)$, with $\mu$ the renormalization scale. As
indicated, quark mass terms are neglected here.
Their effect at \nlo\ has been studied in \citere{Harlander:2016vzb} and it was found to be negligible.
In this case, the
$e_i$ depend only on the product $z\equiv \mu^2t$. For reasons that will become
clear shortly, it is convenient to parameterize this dependence in
terms of the variable
\begin{equation}
  \begin{split}
    L(z) \equiv \ln(2z) + \EulerGamma\,.
  \end{split}
\end{equation}
A
natural choice for the renormalization scale is defined by the inverse
``smearing radius'' $q_8\equiv 1/\sqrt{8t}$ of the gradient
flow. However, from explicit higher order calculations, it was found
that the slightly larger value
\begin{equation}
  \begin{split}
    \mu_0 = \frac{e^{-\EulerGamma/2}}{\sqrt{2t}} \approx 1.5\,q_8\,,
    \label{eq:mu0}
  \end{split}
\end{equation}
corresponding to $L(\mu_0^2t)=0$, improves the stability of the
perturbative corrections\,\cite{Harlander:2018zpi} (see also
\citere{Harlander:2016vzb}). This was corroborated by the behavior of
the $t\to 0$ extrapolation for thermodynamical quantities evaluated in
\citere{Iritani:2018idk} using the \two-loop result of
\citere{Harlander:2018zpi}.

We thus write the coefficients of \eqn{eq:et} as
\begin{equation}
  \begin{split}
    e_0(z) &= e_{0,0}\,,\qquad
    e_1(z) = e_{1,0} + \beta_0\,L(z)\,,\\
    e_2(z) &= e_{2,0} + (2\beta_0\,e_{1,0}+\beta_1)\,L(z) +
    \beta_0^2\,L^2(z)\,,
    \label{eq:efull}
  \end{split}
\end{equation}
with $\beta_0$, $\beta_1$ from \eqn{eq:anombg}.  For the coefficients
$e_{i,j}$, we find
\begin{equation}
  \begin{split}
    e_{0,0} =&\ 1\,,\qquad
    e_{1,0} = \left(\frac{52}{9} + \frac{22}{3} \ln 2 - 3 \ln 3\right)
    \CA - \frac{8}{9} \TF\\
    e_{2,0} =&\ 27.9786\, \CA^2  - (31.5652\ldots)\, \TF\CA
    + \left(16 \zeta(3) - \frac{43}{3}\right)
    \TF\CF + \left(\frac{8\pi^2}{27} -
    \frac{80}{81}\right) \TF^2\,,
    \label{eq:econst}
  \end{split}
\end{equation}
where $\zeta(z)$ is Riemann's $\zeta$ function with $\zeta(3) =
1.20206\ldots$. The three dots in the coefficient of $\TF\CA$ indicate
that we were able to obtain an expression in analytical form. It can be
found in Appendix\,\ref{app:analytic}, \eqn{eq:e20catf}.  Only four
decimal places of the numerical result for the $\CA^2$ coefficient are
displayed here, while our estimate of the numerical accuracy, obtained by
propagating the uncertainty of the individual integrals to the final
result, is about six digits beyond that. This estimate matches the
observed differences between the numerical and analytical results in the
cases where the latter have been computed.  The same statements hold for
the other observables listed below.  The \nlo\ coefficient $e_1$ was
first evaluated in \citere{Luscher:2010iy}.  Setting $\mu=q_8$, one
finds that the \nnlo\ result agrees with \citere{Harlander:2016vzb} at the
sub-percent level.\footnote{To be precise, all color coefficients of
  \citere{Harlander:2016vzb} are compatible with our new calculation,
  except for the $\CA^2$ term, for which ``only'' the first three digits
  agree, corresponding to an overly optimistic uncertainty estimate of
  \citere{Harlander:2016vzb} for that term. The effect for any practical
  application should be irrelevant.}

The logarithmic terms $e_{n,k}L^k(z)$ obey the all-order recursion
formula ($1\leq k\leq n$)
\begin{equation}
  \begin{split}
    k\,e_{n,k} = \sum_{l=0}^{n-1}(n-l)e_{n-l-1,k-1}\beta_l\,,
  \end{split}
\end{equation}
which follows from renormalization-group invariance, i.e.
\begin{equation}
  \begin{split}
    \mu\dderiv{}{\mu}\langle E(t)\rangle = 0\,,\qquad
    \mu^2\dderiv{}{\mu^2}\alphas = \alphas\beta(\alphas) =
    -\frac{\alpha^2_s}{4\pi}\left(\beta_0 + \frac{\alphas}{4\pi}\beta_1
    + \ldots\right)\,.
    \label{eq:rgee}
  \end{split}
\end{equation}
This provides a welcome check of our calculation.  The residual
dependence on $\mu$ can be used as probe of the perturbative behavior.
We refer to \citere{Harlander:2016vzb} for such a study.

Since $\langle E(t) \rangle$ is also gauge independent, we are free to
choose Feynman gauge for the \qcd\ gauge parameter ($\xi=1$) and $\kappa = 1$ for the gradient-flow gauge parameter
for the evaluation of the results
presented above, which facilitates the actual calculation
significantly. However, we also check gauge invariance explicitly by
evaluating the terms with the highest power in the gauge parameter
$\xi$, i.e.\ $\xi^4$, and show that its coefficient vanishes. Indeed,
this already happens algebraically after the reduction to master
integrals, even before their numerical values are inserted.  Keeping the
gauge parameter fully general unfortunately leads to intermediate
expressions which exceed our available computing resources.

\subsection{Results for the quark condensate at three loops}

To obtain the quark condensate with ringed fields we first consider the
conversion factor $\zeta_\chi(t,\mu)$ from \msbar\ renormalized to
ringed quark fields as defined in \eqn{eq:zetachi}. Applying the methods
described in \sct{sec:pert}, we can calculate the diagrams contributing
to the Green's function $\langle R(t)\rangle$ for massless quarks, which
eventually leads to the result
\begin{equation}
	\begin{split}
	  \zeta_\chi(t,\mu) = 1 &+ \frac{\alphas}{4\pi} \left(
          \frac{\gamma_{\chi, 0}}{2} L(\mu^2t) - 3 \CF \ln 3 - 4 \CF \ln
          2 \right) \\ &+ \left(\frac{\alphas}{4\pi}\right)^2 \Bigg\{
          \frac{\gamma_{\chi, 0}}{4} \left( \beta_0 +
          \frac{\gamma_{\chi, 0}}{2} \right) L^2 (\mu^2t) + \Big[
            \frac{\gamma_{\chi, 1}}{2} - \frac{\gamma_{\chi, 0}}{2}
            \left( \beta_0 + \frac{\gamma_{\chi, 0}}{2} \right) \ln 3
            \\ & \qquad \qquad - \frac{2}{3} \gamma_{\chi, 0} \left(
            \beta_0 + \frac{\gamma_{\chi, 0}}{2}\right) \ln 2 \Big]
          L(\mu^2t) + C_2 \Bigg\} + \order{\alphas^3,m} \,.
                \label{eq:zfchi}
	\end{split}
\end{equation}
The \nlo\ result has been obtained in \citere{Makino:2014taa}, and the
$\mu$-dependent \nnlo\ terms were derived in
\citere{Harlander:2018zpi}. That paper also used a preliminary result
for the coefficient $C_2$, derived for the first time in the current
paper. Keeping four decimal places for the individual color coefficients,
we find
\begin{equation}
  C_2 = -23.7947 \, \CA \CF + 30.3914 \, \CF^2 -(3.9226\ldots) \CF
  \TF\,,
  \label{eq:c2}
\end{equation}
where again, as indicated by the three dots, the coefficient of $\CF\TF$
was obtained in analytical form and is given in
Appendix\,\ref{app:analytic}, \eqn{eq:c2cftf}.

Let us now turn to the quark condensate $\langle S(t)\rangle$. Since it
vanishes in the chiral limit $m_f=0$, we compute the leading coefficient
of the expansion in the quark masses $m_f$,
\begin{align}
  \left\langle \zeta_\chi S(t) \right\rangle \equiv \langle
  \mathring{S}(t) \rangle = \sum_{f=1}^{\NF} s_f(t) + \mathcal{O}(m^2)\,,\qquad
  s_f(t) \equiv m_f\dderiv{\langle
    \mathring{S}(t)\rangle}{m_f}\bigg|_{m=0}\,,
  \label{eq:sring}
\end{align}
where it is understood that \textit{all} quark masses are set to zero
after taking the derivative.  This expansion can be applied before the
loop and flow-time integrations are carried out, which means that, in
the quark propagators, we write
\begin{equation}
  \begin{split}
    \frac{1}{\mathrm{i}\slashed{q}+m_f} =
    \frac{1}{\mathrm{i}\slashed{q}} + \frac{m_f}{q^2} +
    \mathcal{O}(m_f^2)\,,
    \label{eq:massexp}
  \end{split}
\end{equation}
and keep only the leading linear term in $m$ of the integrand. The resulting
integrals are then again of the form shown in \eqn{eq:intrep}. We thus
write
\begin{equation}
  s_f(t) = - \frac{\NC m_f}{8\pi^2t} \left[s_0 + \frac{\alphas}{4\pi} s_1 +
    \left(\frac{\alphas}{4\pi}\right)^2 s_2 +
    \mathcal{O}(\alphas^3)\right] + \mathcal{O}(m^2) \,.
\end{equation}
Renormalizing the quark masses in the \msbar\ scheme according to
\eqn{eq:renorm}, we find
\begin{equation}
  \begin{split}
    s_0(z) &= s_{0,0}\,,\qquad
    s_1(z) = s_{1,0} + \frac{\gamma_{m,0}}{2}\,L(z)\,,\\
    s_2(z) &= s_{2,0} + \frac{1}{2}\left[(\gamma_{m,0}+2\beta_0)\,s_{1,0} +
      \gamma_{m,1}\right]L(z) +
    \frac{1}{8}(\gamma_{m,0}+2\beta_0)\gamma_{m,0}\,L^2(z)\,,
    \label{eq:sfull}
  \end{split}
\end{equation}
with $z=\mu^2t$ and the $\gamma_{m,i}$ and $\beta_i$ from \eqn{eq:anombg}.
The non-logarithmic terms are given by
\begin{equation}
  \begin{split}
    s_{0,0} &= 1\,,\qquad s_{1,0} = (4 + 4 \ln 2 - 3 \ln 3) \CF \\
    s_{2,0} &= 19.6422\, \CF^2 + 41.3897\, \CF \CA - (15.7975\dots)\, \TF \CF\,,
    \label{eq:sconst}
  \end{split}
\end{equation}
where again the $\TF\CF$ term is known analytically, as indicated by the
three dots. It can be found in Appendix\,\ref{app:analytic},
\eqn{eq:s20cftf}.  The \nlo\ term has been obtained in
\citere{Makino:2018rys}, the \nnlo\ term is new. Again, only four decimal
places are displayed and our uncertainty estimate for the numerical
integration is several digits beyond that.

\begin{figure}
  \begin{center}
    \begin{tabular}{cc}
      \includegraphics[width=.45\textwidth]{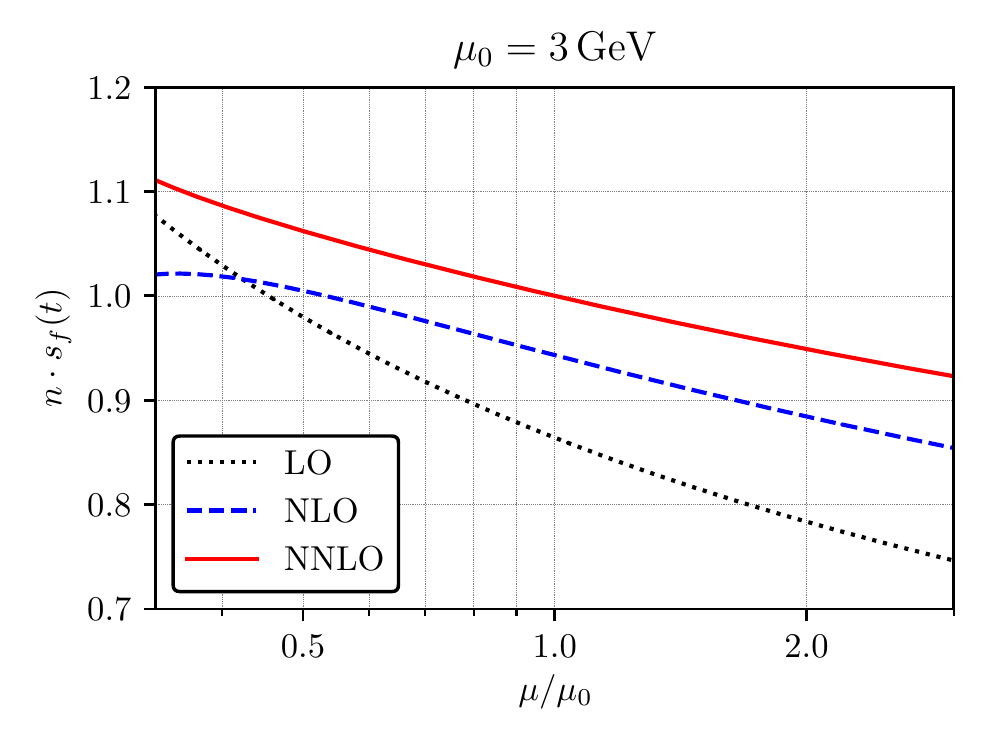} &
      \includegraphics[width=.45\textwidth]{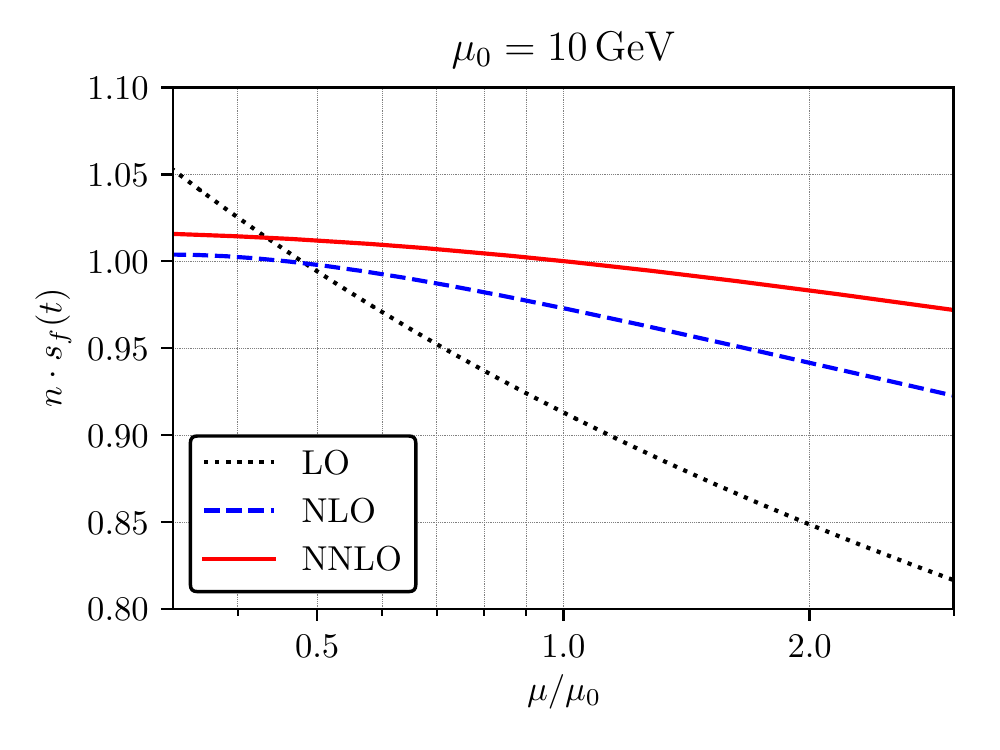}\\
      \hspace*{2em}(a) & \hspace*{2em}(b)\\
      \multicolumn{2}{c}{\includegraphics[width=.5\textwidth]{%
          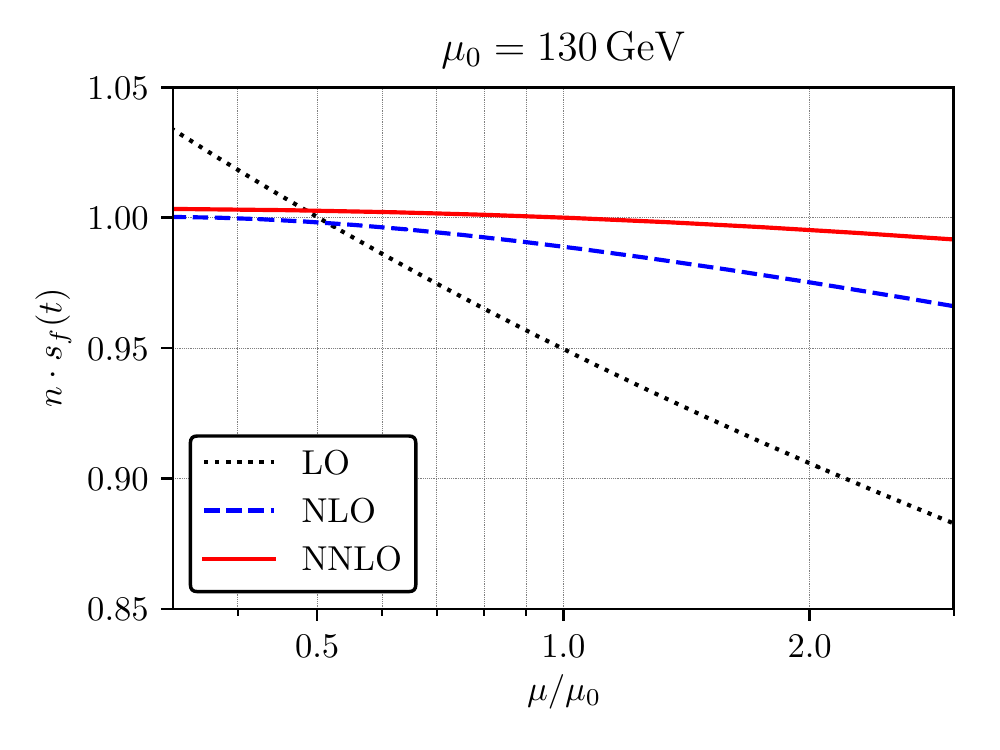}}\\
      \multicolumn{2}{c}{\hspace*{2em}(c)}
    \end{tabular}
  \end{center}
	\caption{Renormalization scale dependence of $s_f(t)$, defined
          in \eqn{eq:sring}, for three different values of the central
          scale
          $\mu_0=e^{-\EulerGamma/2}/\sqrt{2t}$. (a)~$\mu_0=3$\,GeV,
          corresponding to $t=(0.03\,\mathrm{fm})^2$, where we use $\NF = 3$
          and $\alpha_\mathrm{s}^{(3)}(3\,\mathrm{GeV}) =
          0.248$;
          (b)~$\mu_0=10$\,GeV ($t=(0.01\,\mathrm{fm})^2$), where $\NF = 5$ and
          $\alpha_\mathrm{s}^{(5)}(10\,\mathrm{GeV}) = 0.178$;
          (c)~$\mu_0=130$\,GeV ($t=(8\cdot
          10^{-4}\,\mathrm{fm})^2$), where $\NF = 5$ and
          $\alpha_\mathrm{s}^{(5)}(130\,\mathrm{GeV}) = 0.112$.
          The
          running of $\alphas(\mu)$ and $m_f(\mu)$ is evaluated with the
          help of
          \texttt{RunDec}\,\cite{Chetyrkin:2000yt,Herren:2017osy} at the
          corresponding loop order. All curves are normalized to the
          \nnlo\ result at $\mu=\mu_0$.}
	\label{fig:S_plot}
\end{figure}

The logarithmic terms $s_{n,k}L^k(z)$ obey the all-order recursion formula ($1\leq k\leq n$)
\begin{equation}
  \begin{split}
k\,s_{n,k} = \sum_{l=0}^{n-1}\left[\frac{\gamma_{n,l}}{2} +
  (n-l-1)\beta_l\right]s_{n-l-1,k-1}\,,
  \end{split}
\end{equation}
which follows from renormalization group
invariance, i.e.
\begin{equation}
  \begin{split}
    \mu\dderiv{}{\mu} s_f(t) = 0\,,\qquad
    \mu\dderiv{}{\mu}m_f = m_f\gamma_m(\alphas) =
    - m_f\frac{\alphas}{4\pi}\left(\gamma_{m,0} +
    \frac{\alphas}{4\pi}\gamma_{m,1} + \ldots\right)\,,
  \end{split}
\end{equation}
and the renormalization group equation for $\alphas$ given in
\eqn{eq:rgee}. This again provides a welcome check of our calculation.
The residual dependence on $\mu$ can be used as probe of the
perturbative behavior. It is displayed in \fig{fig:S_plot} for three
values of the central energy scale $\mu_0=(2te^{\EulerGamma})^{-1/2}$,
see \eqn{eq:mu0}. We observe a qualitatively similar behavior as was
found for $\langle t^2E(t)\rangle$ in \citere{Harlander:2016vzb}. While
for large $\mu_0$, the reduction of the renormalization scale with
increasing perturbative order is quite significant, this improvement
reduces as $\mu_0$ gets closer to the non-perturbative regime. Taking
the variation of $\mu$ by a factor of two around the central scale as an
estimate of the uncertainty due to missing higher orders, one finds that
they overlap between successive perturbative orders for all three values
of $\mu_0$. However, the trend of the curves does not explicitly
support $\mu_0$ as the central scale in this case, but seems to favor a
slightly smaller scale.

Another important consistency check is that $\langle S(t) \rangle$ and
$\langle R(t) \rangle$ can be renormalized with the same flowed-quark
field renormalization constant $Z_\chi$ as obtained in
\citere{Harlander:2018zpi}.  $\langle S(t) \rangle$ and $\langle
R(t)\rangle$ are also gauge independent.  For the calculation of the
full \nnlo\ result, we again chose Feynman gauge ($\xi=1$) and
$\kappa=1$.  We check gauge invariance explicitly by evaluating the
terms with the highest power in the gauge parameter $\xi$, which is
$\xi^3$ for $\langle S(t) \rangle$ and $\langle R(t)\rangle$, and show
that their coefficients vanish.  This happens here also immediately
after the reduction to master integrals.

\section{Gradient-flow coupling and mass}
\label{sec:couplingandmass}

\subsection{Gradient-flow coupling}

The renormalization group invariance of $\langle E(t)\rangle$ allows us
to fix a scale $\mu=\sqrt{\rho/t}$ with constant $\rho\in\mathbb{R}$,
and its proportionality to $\alphas$ suggests to define a
``gradient-flow coupling''
\begin{equation}
  \begin{split}
    \alphagf(\mu) \equiv \frac{32\pi\rho^2}{3\NA\mu^4}\,\langle
    E(\rho/\mu^2)\rangle \equiv
    \alphas(\mu)\sum_{n=0}^\infty\left(\frac{\alphas(\mu)}{4\pi}\right)^n
    e_{n}(\rho)\,.
  \end{split}
\end{equation}
For the usual definition of the gradient-flow
coupling\,\cite{Luscher:2010iy}, $\rho$ is set to $1/8=0.125$, while
$\rho=1/(2e^{\EulerGamma})=0.281\ldots$ appears to be more suitable from
a perturbative point of view (cf.\ \eqn{eq:mu0}).

Perturbatively solving \eqn{eq:et} for $\alphas$ results in
\begin{equation}
  \begin{split}
    \alphas = \alphagf\left[ 1 - e_{1}(\rho)\frac{\alphagf}{4\pi} +
      (2e_{1}^2(\rho)-e_{2}(\rho))\left(\frac{\alphagf}{4\pi}\right)^2 +
      \ldots\right]\,,
  \end{split}
\end{equation}
with the coefficients $e_{n}(\rho)$ given in \eqn{eq:efull}. Such a
transformation between mass-indepen\-dent renormalization schemes leaves
the first two coefficients of the $\beta$ function invariant, i.e.\
\begin{equation}
  \begin{split}
    \mu^2\dderiv{}{\mu^2}\alphagf(\mu) &= \alphagf(\mu)\betagf{}(\alphagf)\,,\\
    \betagf{}(\alphagf) = -\sum_{n=0}^\infty\betagf{n}
    \left(\frac{\alphagf}{4\pi}\right)^n
    &= - \frac{\alphagf}{4\pi}\beta_0
     - \left(\frac{\alphagf}{4\pi}\right)^2\beta_1
     -\sum_{n=2}^\infty\betagf{n}
     \left(\frac{\alphagf}{4\pi}\right)^n\,,
  \end{split}
\end{equation}
with $\beta_0$ and $\beta_1$ from \eqn{eq:anombg}. The third
coefficient is given by
\begin{equation}
  \begin{split}
    \betagf{2} = \beta_2 - e_{1}(\rho)\,\beta_1 + (e_{2}(\rho)
    - e_{1}^2(\rho))\,\beta_0\,,
  \end{split}
\end{equation}
with the \msbar\ coefficient $\beta_2$ which we quote here for the
SU(3) gauge group:
\begin{equation}
  \begin{split}
    \beta_2 = \frac{2857}{2} - \frac{5033}{18}\NF + \frac{325}{54}\NF^2\,.
  \end{split}
\end{equation}

\subsection{Gradient-flow mass}

Similarly, we may define a gradient-flow quark mass as
\begin{equation}
  \begin{split}
    \massgf_f(\mu) = -\frac{8\pi^2 \rho}{\NC\mu^2}m_f\dderiv{\langle
      \mathring{S}(\rho/\mu^2)\rangle}{m_f}\Big|_{m=0} = m_f(\mu)\sum_{n=0}^\infty
    \hat{s}_{n}(\rho)\left(\frac{\alphagf(\mu)}{4\pi}\right)^n\,,
  \end{split}
\end{equation}
where
\begin{equation}
  \begin{split}
    \hat{s}_{0}(\rho) = s_{0}(\rho)\,,\qquad
    \hat{s}_{1}(\rho) = s_{1}(\rho)\,,\qquad
    \hat{s}_{2}(\rho) = s_{2}(\rho) - s_{1}(\rho)\, e_{1}(\rho)\,,
  \end{split}
\end{equation}
with the $s_{n}$ and $e_{1}$ given in \eqs{eq:sfull} and
\noeqn{eq:efull}. The inverse relation is
\begin{equation}
  \begin{split}
    m_f = \massgf_f\left[1-\hat{s}_{1}(\rho)\frac{\alphagf}{4\pi} +
      (\hat{s}_{1}^2(\rho)-\hat{s}_{2}(\rho))
      \left(\frac{\alphagf}{4\pi}\right)^2
      + \ldots\right]\,.
  \end{split}
\end{equation}
The new mass obeys the renormalization group equation
\begin{equation}
  \begin{split}
    \mu\dderiv{}{\mu}\massgf_f(\mu) = \gammagf_m(\alphagf)\,\massgf_f(\mu)
   \equiv - \massgf_f(\mu)\,\sum_{n=0}^\infty
    \left(\frac{\alphagf(\mu)}{4\pi}\right)^{n+1}\gammagf_{m,n}
    \label{eq:rgme}
  \end{split}
\end{equation}
where
\begin{equation}
  \begin{split}
    \gammagf_{m,0} &= \gamma_{m,0}\,,\qquad
    \gammagf_{m,1} = \gamma_{m,1}
    - \gamma_{m,0}\, e_{1}(\rho)
    + 2\,\beta_0\,\hat{s}_{1}(\rho)
    \,,\\
    \gammagf_{m,2} &=
    \gamma_{m,2}
    - 2\,\gamma_{m,1}\,e_{1}(\rho)
    + \gamma_{m,0}\,\left[2\,e_{1}^2(\rho) - e_{2}(\rho)\right]\\&\qquad
    + 2\,\beta_1\hat{s}_{1}(\rho)
    + 2\,\beta_0\,\left(2\,\hat{s}_{2}(\rho)-\hat{s}^2_1(\rho)\right)\,,
  \end{split}
\end{equation}
with $\beta_0$, $\beta_1$, $\gamma_{m,0}$, $\gamma_{m,1}$ given in
\eqn{eq:anombg}, and $\gamma_{m,2}$ the \three-loop \msbar\ coefficient
which we quote for the SU(3) gauge group:
\begin{equation}
  \begin{split}
    \gamma_{m,2} = 2498 -
    \left(\frac{4432}{27}+\frac{320}{3}\zeta(3)\right) \NF -
    \frac{280}{81}\,\NF^2\,.
  \end{split}
\end{equation}

\section{Conclusions}
\label{sec:conclusions}

A framework for the calculation of correlation functions in the
perturbative gradient-flow formalism using tools and techniques from
standard perturbative \abbrev{QCD} calculations has been presented. It
employs a Feynman-diagrammatic approach to the gradient-flow formalism,
based on the five-dimensional field-theoretical formulation of
\citere{Luscher:2011bx}. We implemented the corresponding \qcd\ Feynman
rules within the gradient-flow formalism as well as those of the
composite operators relevant for the quantities considered in this
paper, and automated the reduction and calculation of algebraic
expressions and loop integrals to a large extent. An important
ingredient of our setup is the reduction to master integrals with the
help of integration-by-parts identities that also involve the flow-time
variables. Except for a few cases where an analytic result was derived,
the master integrals were evaluated numerically.

To demonstrate the viability of this setup, we reproduced and improved
the three-loop result for the gluon condensate of an earlier
calculation, which was based on explicit field-theoretic calculations at
the level of Wick contractions\,\cite{Harlander:2016vzb}.

Using this setup, we then evaluated the three-loop approximations to the
quark condensate and the conversion factor from \msbar{} renormalized to
ringed quark fields, both of which are new results. Checks on gauge
parameter independence, renormalization group invariance, as well as
alternative calculational approaches (with and without reduction to
master integrals) convinced us of the correctness of these results.

The gluon and the quark condensate lend themselves for straightforward
definitions of a gradient-flow gauge coupling and a gradient-flow
quark mass, for which we derive the three-loop matching relations to the
$\msbar$ scheme. Provided that sufficiently accurate lattice results for
these quantities become available, the conversion factors evaluated in
this paper should allow for precise first-principle determinations of
the gauge coupling,\footnote{First steps in this direction have been
  taken in \citere{Lambrou:2016uet}.} and possibly also quark masses.

While focusing on the calculation of vacuum matrix elements in this
paper, our setup should be applicable in principle to more general
problems which may involve a larger number of scales, for example.

\section*{Acknowledgments}

We would like to thank Lucius Bushnaq and Yannick Kluth for their help
in checking the Feynman rules and the general setup, Jonas Klappert for
his support in using \texttt{FireFly}, Johann Usovitsch for his support
in using \texttt{Kira}, Alexander Voigt for general technical
discussions, and Mauro Papinutto for pointing out a typo in an earlier
version of this manuscript.  This work was supported by \textit{Deutsche
  Forschungsgemeinschaft (DFG)} through project
\href{http://gepris.dfg.de/gepris/projekt/386986591}{HA 2990/9-1}, and
by the U.S.\ Department of Energy under award No.\ DE-SC0008347.  This
document was prepared using the resources of the Fermi National
Accelerator Laboratory (Fermilab), a U.S. Department of Energy, Office
of Science, HEP User Facility. Fermilab is managed by Fermi Research
Alliance, LLC (FRA), acting under Contract No.\ DE-AC02-07CH11359.  Some
of the calculations were performed with computing resources granted by
\abbrev{RWTH} Aachen University under project \textit{rwth0244}.  All
Feynman diagrams in this paper were produced using
Ti\textit{k}Z-Feynman\,\cite{Ellis:2016jkw}.

\appendix

\section{Feynman rules}
\label{sec:frules}

In this section we present all the Feynman rules required for our
calculation.  All momenta are considered to be outgoing.  $\xi$ is the
usual \qcd\ gauge parameter and $\kappa$ the additional gradient-flow
gauge parameter introduced in the flow equations \noeqn{eq:flow}.  Lines
with a solid arrow are flow lines and lines with a dashed arrow can be
both flow lines or flowed propagators. The symbols $p,q,r$ denote
Euclidean four-momenta, while $s,t$ are real-valued flow-time variables.

\textbf{Propagators and flow lines}
\begin{itemize}
\item gluon propagator:
\begin{flalign}
  & \raisebox{-0.5em}{\includegraphics{dias/gluon-propagator.pdf}}
  & = &
  \begin{split}
    \delta^{ab} \frac{1}{p^2} \bigg( & \left( \delta_{\mu \nu} - \frac{p_\mu p_\nu}{p^2} \right) e^{-(t+s)p^2} \\
    & + \xi \frac{p_\mu p_\nu}{p^2} e^{-\kappa(t+s)p^2} \bigg)
  \end{split}
  & &
\end{flalign}

\item (anti)quark propagator:
\begin{flalign}
  & \raisebox{-0.5em}{\includegraphics{dias/quark-propagator.pdf}}
  & = &
  \begin{split}
    \delta_{ij} \frac{(-\mathrm{i}\slashed{p}+m)_{\alpha\beta}}{p^2+m^2} \, e^{-(t+s)p^2}
  \end{split}
  & &
\end{flalign}

\item ghost propagator:
\begin{flalign}
  & \raisebox{-0.5em}{\includegraphics{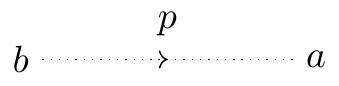}}
  & = &
  \begin{split}
    \delta^{ab} \frac{1}{p^2}
  \end{split}
  & &
\end{flalign}

\item gluon flow line:
\begin{flalign}
  & \raisebox{-1.2em}{\includegraphics{dias/gluon-flowline.pdf}}
  & = &
  \begin{split}
    \delta^{ab} \, \theta(t-s) \bigg( & \left( \delta_{\mu \nu} - \frac{p_\mu p_\nu}{p^2} \right) e^{-(t-s)p^2} \\
    &+ \frac{p_\mu p_\nu}{p^2} \, e^{-\kappa(t-s)p^2} \bigg)
  \end{split}
  & &
\end{flalign}

\item quark flow line:
\begin{flalign}
  & \raisebox{-1.2em}{\includegraphics{dias/quark-flowline.pdf}}
  & = &
  \begin{split}
    \delta^{ij} \delta_{\alpha\beta} \, \theta(t-s) \, e^{-(t-s)p^2}
  \end{split}
  & &
\end{flalign}

\item antiquark flow line:
\begin{flalign}
  & \raisebox{-1.2em}{\includegraphics{dias/antiquark-flowline.pdf}}
  & = &
  \begin{split}
    \delta^{ij} \delta_{\alpha\beta} \, \theta(t-s) \, e^{-(t-s)p^2}
  \end{split}
  & &
\end{flalign}
\end{itemize}

\textbf{Vertices}
\begin{itemize}
\item three-gluon vertex:
\begin{flalign}
  &
  \begin{gathered}
    \includegraphics{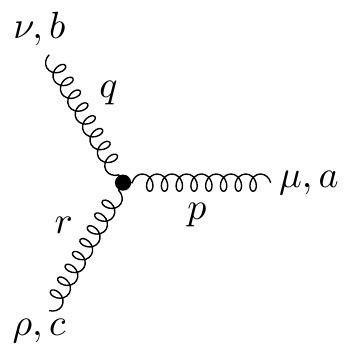}
  \end{gathered}
  & = &
  \begin{split}
    \mathrm{i} g f^{abc} \, \big( \delta_{\mu \nu} (p-q)_\rho + \delta_{\nu\rho} (q-r)_\mu
    + \delta_{\rho\mu} (r-p)_\nu \big)
  \end{split}
  & &
  \label{fr:3gluon}
\end{flalign}

\item four-gluon vertex:
\begin{flalign}
  &
  \begin{gathered}
    \includegraphics{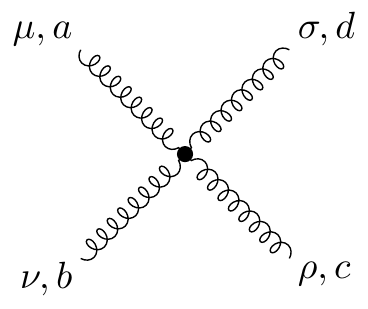}
  \end{gathered}
  & = &
  \begin{split}
    -g^2 \, \big( & f^{abe} f^{cde} (\delta_{\mu\rho}\delta_{\nu\sigma} - \delta_{\mu\sigma}\delta_{\nu\rho} ) \\
    & + f^{ace} f^{bde} (\delta_{\mu\nu}\delta_{\rho\sigma} - \delta_{\mu\sigma}\delta_{\nu\rho} ) \\
    & + f^{ade} f^{bce} (\delta_{\mu\nu}\delta_{\rho\sigma} - \delta_{\mu\rho}\delta_{\nu\sigma} ) \big)
  \end{split}
  & &
\label{fr:4gluon}
\end{flalign}

\item ghost-gluon vertex:
\begin{flalign}
  &
  \begin{gathered}
    \includegraphics{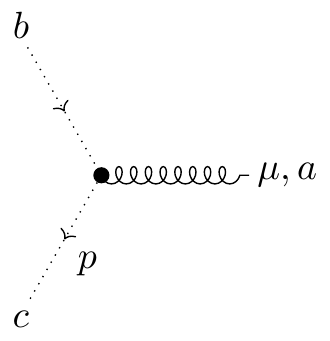}
  \end{gathered}
  & = &
  \begin{split}
    - \mathrm{i} g f^{abc} \, p_\mu
  \end{split}
  & &
\end{flalign}

\item quark-gluon vertex:
\begin{flalign}
  &
  \begin{gathered}
    \includegraphics{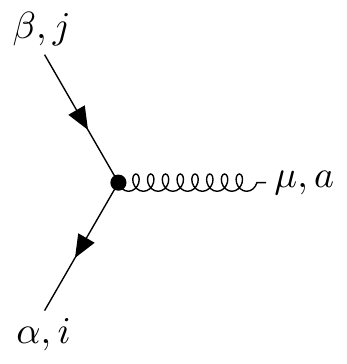}
  \end{gathered}
  & = &
  \begin{split}
    g \, (T^a)_{ij} \, \big( \gamma_\mu \big)_{\alpha\beta}
  \end{split}
  & &
\end{flalign}

\item two-plus-one gluon flow vertex:
\begin{flalign}
  &
  \begin{gathered}
    \includegraphics{dias/ggg-flow-vertex.pdf}
  \end{gathered}
  & = &
  \begin{split}
    -\mathrm{i} g f^{abc} \int_0^\infty \mathrm{d}s \, \big( & \delta_{\nu\rho} (r-q)_\mu + 2 \delta_{\mu\nu} q_\rho - 2 \delta_{\mu\rho} r_\nu \\
    & + (\kappa-1) (\delta_{\mu\rho} q_\nu  - \delta_{\mu\nu} r_\rho) \big) \\
  \end{split}
  & &
\end{flalign}

\item three-plus-one gluon flow vertex:
\begin{flalign}
  &
  \begin{gathered}
    \includegraphics{dias/gggg-flow-vertex.pdf}
  \end{gathered}
  & = &
  \begin{split}
    -g^2 \int_0^\infty \mathrm{d}s \, \big( & f^{abe} f^{cde} (\delta_{\mu\rho}\delta_{\nu\sigma} - \delta_{\mu\sigma}\delta_{\nu\rho} ) \\
    & + f^{ace} f^{bde} (\delta_{\mu\nu}\delta_{\rho\sigma} - \delta_{\mu\sigma}\delta_{\nu\rho} ) \\
    & + f^{ade} f^{bce} (\delta_{\mu\nu}\delta_{\rho\sigma} - \delta_{\mu\rho}\delta_{\nu\sigma} ) \big)
  \end{split}
  & &
\end{flalign}

\item quark-one-gluon flow vertex:
\begin{flalign}
  &
  \begin{gathered}
    \includegraphics{dias/qqg-flow-vertex.pdf}
  \end{gathered}
  & = &
  \begin{split}
    \mathrm{i} g \, \delta_{\alpha\beta} \big( T^a \big)_{ij} \int_0^\infty \mathrm{d} s \, \big( 2 p_\mu + (1-\kappa) q_\mu \big)
  \end{split}
  & &
\end{flalign}

\item antiquark-one-gluon flow vertex:
\begin{flalign}
  &
  \begin{gathered}
    \includegraphics{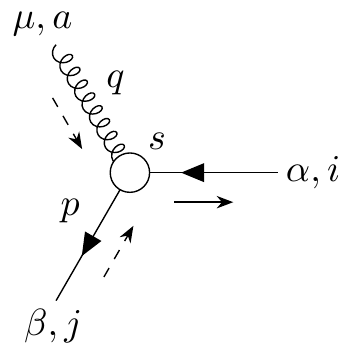}
  \end{gathered}
  & = &
  \begin{split}
    - \mathrm{i} g \, \delta_{\alpha\beta} \big( T^a \big)_{ji} \int_0^\infty \mathrm{d} s \, \big( 2 p_\mu + (1-\kappa) q_\mu \big)
  \end{split}
  & &
\end{flalign}

\item quark-two-gluons flow vertex:
\begin{flalign}
  &
  \begin{gathered}
    \includegraphics{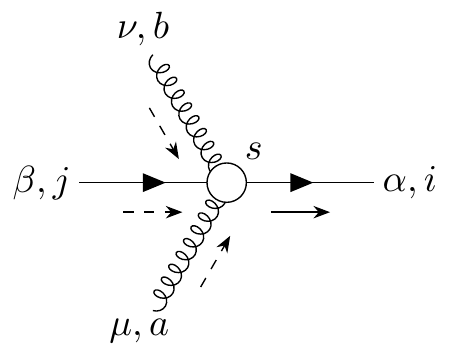}
  \end{gathered}
  & = &
  \begin{split}
    g^2 \delta_{\alpha\beta} \delta_{\mu\nu} \left\{ T^a,T^b \right\}_{ij} \int_0^\infty \mathrm{d}s
  \end{split}
  & &
\end{flalign}

\item antiquark-two-gluons flow vertex:
\begin{flalign}
  &
  \begin{gathered}
    \includegraphics{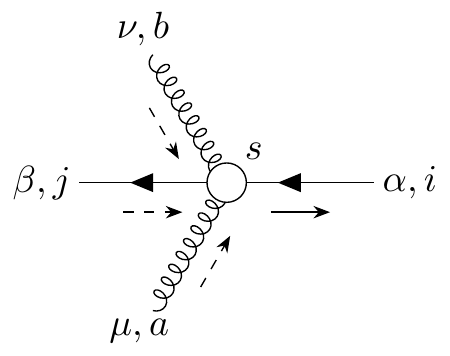}
  \end{gathered}
  & = &
  \begin{split}
    g^2 \delta_{\alpha\beta} \delta_{\mu\nu} \left\{ T^a,T^b \right\}_{ji} \int_0^\infty \mathrm{d}s
  \end{split}
  & &
\end{flalign}
\end{itemize}

Feynman rules for the composite flow-time operators:
\begin{itemize}
\item $G_{\mu\nu}^a(t,x)G_{\mu\nu}^a(t,x)$:
  \begin{itemize}
  \item \two-gluon vertex:
    \begin{flalign}
      &
      \begin{gathered}
        \includegraphics{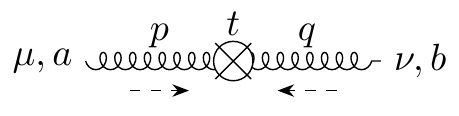}
      \end{gathered}
      & = &
      \begin{split}
        - g^2 \delta^{ab} \big(\delta_{\mu\nu} p \cdot q - p_\mu q_\nu\big)
      \end{split}
      & &
      \label{fr:egg}
    \end{flalign}
  \item \three-gluon vertex:
    \begin{flalign}
      &
      \begin{gathered}
        \includegraphics{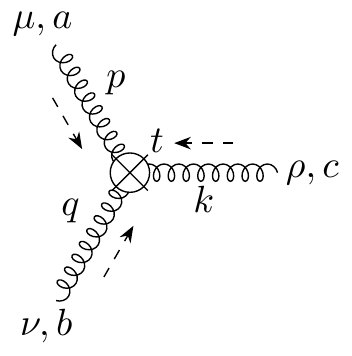}
      \end{gathered}
      & = &
      \begin{split}
        \mathrm{i} g^3 f^{abc} \big( & \delta_{\mu\nu} (q-p)_\rho + \delta_{\nu\rho} (k-q)_\mu \\
        & + \delta_{\mu\rho} (p-k)_\nu \big)
      \end{split}
      & &
      \label{fr:eggg}
    \end{flalign}
  \item \four-gluon vertex:
    \begin{flalign}
      &
      \begin{gathered}
        \includegraphics{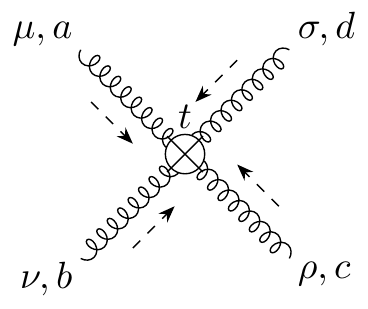}
      \end{gathered}
      & = &
      \begin{split}
        g^4 \big( & f^{abe} f^{cde} (\delta_{\mu\rho}\delta_{\nu\sigma} - \delta_{\mu\sigma}\delta_{\nu\rho} ) \\
        & + f^{ace} f^{bde} (\delta_{\mu\nu}\delta_{\rho\sigma} - \delta_{\mu\sigma}\delta_{\nu\rho} ) \\
        & + f^{ade} f^{bce} (\delta_{\mu\nu}\delta_{\rho\sigma} - \delta_{\mu\rho}\delta_{\nu\sigma} ) \big)
      \end{split}
      & &
      \label{fr:egggg}
    \end{flalign}
  \end{itemize}
\end{itemize}

\begin{itemize}
\item $\bar\chi(t,x)\chi(t,x)$:
    \begin{flalign}
      &
      \begin{gathered}
        \includegraphics{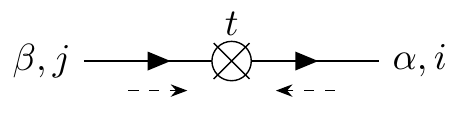}
      \end{gathered}
      & = &
      \begin{split}
        \delta_{ij} \delta_{\alpha\beta}
      \end{split}
      & &
      \label{fr:sqq}
    \end{flalign}
\end{itemize}

\begin{itemize}
  \item $\bar\chi(t,x)\overleftrightarrow{\mathcal{D}}^\mathrm{F}\chi(t,x)$:
    \begin{itemize}
    \item quark-antiquark vertex:
    \begin{flalign}
      &
      \begin{gathered}
        \includegraphics{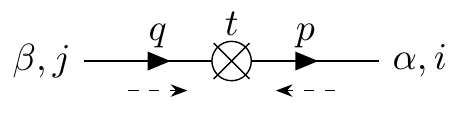}
      \end{gathered}
      & = &
      \begin{split}
        \mathrm{i} \delta^{ij} (\slashed p - \slashed q)_{\alpha\beta}
      \end{split}
      & &
      \label{fr:rqq}
    \end{flalign}
    \item quark-antiquark-gluon vertex:
    \begin{flalign}
      &
      \begin{gathered}
        \includegraphics{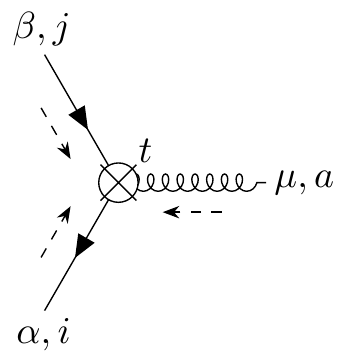}
      \end{gathered}
      & = &
      \begin{split}
        -2 g (T^a)^{ij} (\gamma_\mu)_{\alpha\beta}
      \end{split}
      & &
      \label{fr:rqqg}
    \end{flalign}
    \end{itemize}
\end{itemize}

\textbf{Auxiliary Feynman rules}\\
The following Feynman rules are auxiliary Feynman rules for our implementation as described in \sct{sec:implementation}.
They represent the vertices with four gluons or gluon-flow lines and allow us to factorise the color structure of all Feynman diagrams.

\begin{itemize}
  \item four-gluon vertex:
  \begin{itemize}
    \item $\Sigma$ propagator:
      \begin{flalign}
        &
        \begin{gathered}
          \includegraphics{dias/sigma_propagator.pdf}
        \end{gathered}
        & = &
        \begin{split}
          \delta^{ab}\delta_{\mu\rho}\delta_{\nu\sigma}
        \end{split}
        & &
      \end{flalign}
    \item $\Sigma$-gluon-gluon vertex:
      \begin{flalign}
        &
        \begin{gathered}
          \includegraphics{dias/sigma_vertex.pdf}
        \end{gathered}
        & = &
        \begin{split}
          \frac{\mathrm{i}g}{\sqrt{2}} f^{abc}\left(\delta_{\mu\rho}\delta_{\nu\sigma}-\delta_{\nu\rho}\delta_{\mu\sigma}\right)
        \end{split}
        & &
      \end{flalign}
  \end{itemize}
  \item three-plus-one gluon flow vertex:
    \begin{itemize}
      \item $\Sigma$ flow line:
        \begin{flalign}
          &
          \begin{gathered}
            \includegraphics{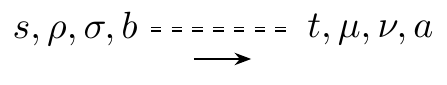}
          \end{gathered}
          & = &
          \begin{split}
            \delta^{ab}\delta_{\mu\rho}\delta_{\nu\sigma} \delta(t-s)
          \end{split}
          & &
        \end{flalign}
      \item outgoing $\Sigma$-gluon-gluon flow vertex:
        \begin{flalign}
          &
          \begin{gathered}
            \includegraphics{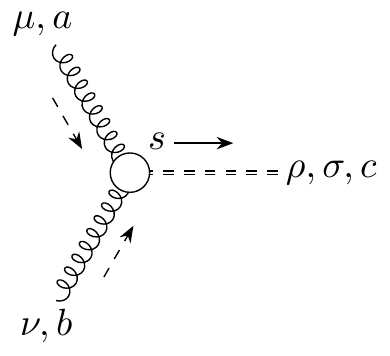}
          \end{gathered}
          & = &
          \begin{split}
            \frac{\mathrm{i}g}{\sqrt{2}} f^{abc}\left(\delta_{\mu\rho}\delta_{\nu\sigma} -\delta_{\nu\rho}\delta_{\mu\sigma}\right) \int_0^\infty \mathrm{d}s
          \end{split}
          & &
        \end{flalign}
      \item incoming $\Sigma$-gluon-gluon flow vertex:
        \begin{flalign}
          &
          \begin{gathered}
            \includegraphics{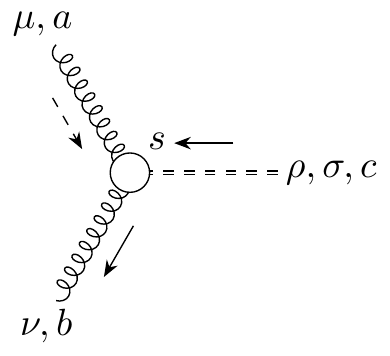}
          \end{gathered}
          & = &
          \begin{split}
            \frac{\mathrm{i}g}{\sqrt{2}} f^{abc}\left(\delta_{\mu\rho}\delta_{\nu\sigma}-\delta_{\nu\rho}\delta_{\mu\sigma}\right) \int_0^\infty \mathrm{d}s
          \end{split}
          & &
        \end{flalign}
    \end{itemize}
    \item \four-gluon vertex from $G_{\mu\nu}^a(t,x)G_{\mu\nu}^a(t,x)$:
      \begin{itemize}
        \item $\Sigma_E$ flow line:
          \begin{flalign}
            &
            \begin{gathered}
              \includegraphics{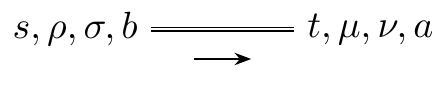}
            \end{gathered}
            & = &
            \begin{split}
              \delta^{ab}\delta_{\mu\rho}\delta_{\nu\sigma} \delta(t-s)
            \end{split}
            & &
          \end{flalign}
        \item outgoing $\Sigma_E$-gluon-gluon vertex:
          \begin{flalign}
            &
            \begin{gathered}
              \includegraphics{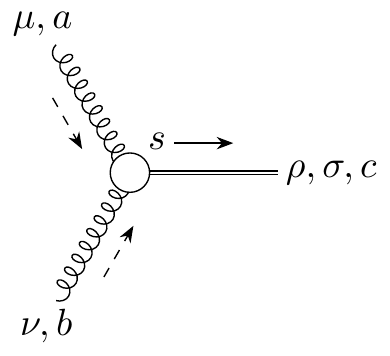}
            \end{gathered}
            & = &
            \begin{split}
              \frac{g^2}{\sqrt{2}} f^{abc}\left(\delta_{\mu\rho}\delta_{\nu\sigma}-\delta_{\nu\rho}\delta_{\mu\sigma}\right) \int_0^\infty \mathrm{d}s
            \end{split}
            & &
          \end{flalign}
        \item incoming $\Sigma_E$-gluon-gluon vertex:
          \begin{flalign}
            &
            \begin{gathered}
              \includegraphics{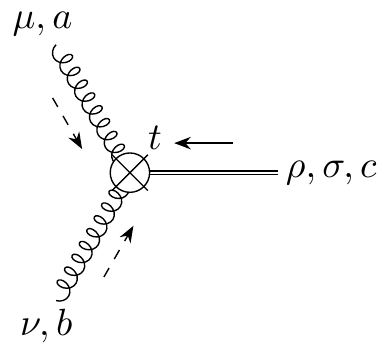}
            \end{gathered}
            & = &
            \begin{split}
              \frac{g^2}{\sqrt{2}} f^{abc}\left(\delta_{\mu\rho}\delta_{\nu\sigma}-\delta_{\nu\rho}\delta_{\mu\sigma}\right)
            \end{split}
            & &
          \end{flalign}
      \end{itemize}
\end{itemize}

\section{Analytical results}\label{app:analytic}

Here, we provide the analytical results for the color coefficients as
far as they are available. For the $\TF\CA$ term of $e_{2,0}$ in
\eqn{eq:econst}, we find
\begin{equation}
  \begin{split}
    -(31.5652\ldots) =&\ \frac{2}{81} \Bigg(-108 \,\text{Li}_3\left(-\frac{1}{3}\right) -792 \,\text{Li}_2\left(\frac{3}{4}\right)+216 \,\text{Li}_2\left(-\frac{1}{3}\right)\\
    &+ 108 \,\text{Li}_{1,2}\left(-3,-\frac{1}{3}\right)-216 \,\text{Li}_{1,2}\left(1,-\frac{1}{3}\right) +216 \,\text{Li}_{2,1}\left(-3,-\frac{1}{3}\right) \\
    &-216 \,\text{Li}_{2,1}\left(1,-\frac{1}{3}\right) +216 \,\text{Li}_{1,1,1}\left(-1,-3,-\frac{1}{3}\right) \\
    &-864 \,\text{Li}_{1,1,1}\left(-1,3,-\frac{1}{3}\right) +216 \,\text{Li}_{1,1,1}\left(1,3,-\frac{1}{3}\right) \\
    &+432 \,\text{Li}_{1,1,1}\left(3,1,-\frac{1}{3}\right) -1134 \zeta (3)+90 \pi ^2-641-54 \ln^3 3 \\
    &+324 \ln^3 2 +324 \ln 2 \ln^2 3+270 \ln^2 3-648 \ln^2 2 \ln 3-2232 \ln^2 2 \\
    &+504 \ln 2 \ln 3-18 \pi ^2 \ln 3+1890\ln 3+108 \pi^2 \ln 2-2148 \ln 2\Bigg) \,,
    \label{eq:e20catf}
  \end{split}
\end{equation}
where
\begin{equation}
  \begin{split}
    \text{Li}_{n_1,\ldots,n_r}(z_1,\ldots,z_r) = \sum_{0<k_1<\cdots<k_r}
    \frac{z_1^{k_1}\cdots z_k^{k_r}}{k_1^{n_1}\cdots k_r^{n_r}}
  \end{split}
\end{equation}
are multiple polylogarithms\,\cite{Goncharov:1998kja,Borwein:1999js}.

The $\CF\TF$ coefficient of $C_2$ in \eqn{eq:c2} reads
\begin{equation}
  \begin{split}
    -(3.9226\ldots) = &\ - \frac{1}{18} \bigg(
    48\,\text{Li}_2\left(\frac{1}{9}\right)-2400\,\text{Li}_2
    \left(\frac{1}{3}\right)+672\,\text{Li}_2\left(\frac{3}{4}\right)-131+46
    \pi ^2 \\ &+960 \ln^2 2 -1068 \ln^2 3 +1888 \ln 2-1032\ln 3+624 \ln
    2 \ln 3 \bigg) \,.
    \label{eq:c2cftf}
  \end{split}
\end{equation}

And finally, the $\CF\TF$ coefficient of $s_{2,0}$ in \eqn{eq:sconst}
reads
\begin{equation}
  \begin{split}
    -(15.7975 \dots) =
    &\ \frac{16}{3} \, \text{Li}_2\left(\frac{1}{9}\right) - \frac{800}{3} \, \text{Li}_2\left(\frac{1}{3}\right) + 104 \, \text{Li}_2\left(\frac{3}{4}\right) -\frac{71}{6}+\frac{23 \pi ^2}{9} \\
    &+ 192 \ln^2 2 - \frac{362}{3} \ln^2 3 +\frac{544}{9} \ln 2-\frac{100}{3} \ln 3 + \frac{56}{3} \ln 2 \ln 3 \,.
    \label{eq:s20cftf}
  \end{split}
\end{equation}

\newcommand{\bibentry}[4]{#1, {\it #2}, #3\ifthenelse{\equal{#4}{}}{}{, }#4.}
\newcommand{\journal}[5]{\href{https://dx.doi.org/#5}{\textit{#1} {\bf #2} (#3) #4}}
\newcommand{\arxiv}[2]{\href{https://arXiv.org/abs/#1}{\texttt{arXiv:#1\,[#2]}}}
\newcommand{\arxivhepph}[1]{\href{https://arXiv.org/abs/hep-ph/#1}{\texttt{hep-ph/#1}}}
\newcommand{\arxivhepth}[1]{\href{https://arXiv.org/abs/hep-th/#1}{\texttt{hep-th/#1}}}
\newcommand{\arxivheplat}[1]{\href{https://arXiv.org/abs/hep-lat/#1}{\texttt{hep-lat/#1}}}
\newcommand{\arxivmathph}[1]{\href{https://arXiv.org/abs/math-ph/#1}{\texttt{math-ph/#1}}}
\newcommand{\arxivmath}[1]{\href{https://arXiv.org/abs/math/#1}{\texttt{math/#1}}}

\IfFileExists{./\jobname_ref.tex}{
  
}{}

\end{document}